\documentclass[3p, twocolumn, times, authoryear]{elsarticle}

\usepackage{amsmath} 
\usepackage{breqn} 

\usepackage{graphicx}
\usepackage{subfigure}

\usepackage[none]{hyphenat}
\usepackage{multibib}
\newcites{paper}{References}
\newcites{appendix}{References}

\bibliographystylepaper{apalike}
\bibliographystyleappendix{apalike}

\begin{document}

\begin{frontmatter}

\title{The thermodynamics of protein aggregation reactions may underpin the enhanced metabolic efficiency associated with heterosis, some balancing selection, and the evolution of ploidy levels}
\author[uga]{B.R. ~Ginn}
\ead{bginn3@gmail.com}

\begin{abstract}
	Identifying the physical basis of heterosis (or ``hybrid vigor'') has remained elusive despite over a hundred years of research on the subject. The three main theories of heterosis are dominance theory, overdominance theory, and epistasis theory. \cite{KacserAndBurns1981} identified the molecular basis of dominance, which has greatly enhanced our understanding of its importance to heterosis. This paper aims to explain how overdominance, and some features of epistasis, can similarly emerge from the molecular dynamics of proteins. Possessing multiple alleles at a gene locus results in the synthesis of different allozymes at reduced concentrations. This in turn reduces the rate at which each allozyme forms soluble oligomers, which are toxic and must be degraded, because allozymes co-aggregate at low efficiencies. The model developed in this paper can explain how heterozygosity impacts the metabolic efficiency of an organism. It can also explain why the viabilities of some inbred lines seem to decline rapidly at high inbreeding coefficients (F $>$ 0.5), which may provide a physical basis for truncation selection for heterozygosity. Finally, the model has implications for the ploidy level of organisms. It can explain why polyploids are frequently found in environments where severe physical stresses promote the formation of soluble oligomers. The model can also explain why complex organisms, which need to synthesize aggregation-prone proteins that contain intrinsically unstructured regions (IURs) and multiple domains because they facilitate complex protein interaction networks (PINs), tend to be diploid while haploidy tends to be restricted to relatively simple organisms.

\end{abstract}


\end{frontmatter}

\section{Introduction} \label{sec:Introduction}

	Heterosis, or ``hybrid vigor'', refers to the superior performance of highly heterozygous individuals relative to less heterozygous individuals on a number of biological metrics (\citealt{LippmanAndZamir2007}; \citealt{HochholdingerAndHoecker2007}; \citealt{BirchlerEtAl2010}). Biologists have been aware of the phenomena for over a hundred years (e.g. \citealt{Darwin1876}), and it has been exploited to improve crop yields substantially over the twentieth century, especially in maize (\citealt{Crow1998}; \citealt{Duvick2001}), but there is still debate over its origin (for review see \citealt{LippmanAndZamir2007}; \citealt{HochholdingerAndHoecker2007}; \citealt{BirchlerEtAl2010}). Three different theories are usually used to explain heterosis: dominance theory (\citealt{Davenport1908}; \citealt{Bruce1910}; \citealt{Jones1917}), overdominance theory (\citealt{Shull1948}; \citealt{East1936}), and epistasis (\citealt{Powers1944}; \citealt{Williams1959}). Dominance theory attributes the benefits of heterozygous genotypes to the masking of recessive deleterious alleles. Proponents of overdominance theory argue that heterozygosity itself can have benefits, even in the absence of deleterious mutations. Proponents of epistasis argue that heterosis comes from positive interactions between multiple gene loci. Currently, dominance theory is the most widely accepted theory of heterosis (\citealt{Crow1998}; \citealt{CharlesworthAndWillis2009}).

	However, dominance theory is unlikely to be the sole explanation for heterosis. Five lines of evidence from the literature for overdominance and epistasis are briefly mentioned here. First, the performance of hybrid rice cannot be explained solely by dominance theory, and may require some combination of overdominant and epistatic interactions (\citealt{LiEtAl2001}; \citealt{ZhouEtAl2012}). Second, breeding experiments performed on polyploid plants appear to indicate that possessing three or more alleles at a gene locus is more beneficial than possessing two, which is difficult to explain with dominance theory alone (\citealt{GrooseEtAl1989}; \citealt{RiddleAndBirchler2008}). Third, the rate of heterozygosity decline that occurs after multiple generations of inbreeding is slower than predicted by dominance theory (\citealt{RumballEtAl1994}; \citealt{DemontisEtAl2009};\citealt{CheloAndTeotonio2012}). The slow decline may reflect balancing selection (selection for overdominant loci) or linkage between the measured genetic markers and deleterious alleles (associative overdominance). Fourth, haplodiploid (\citealt{Henter2003}; \citealt{TortajadaEtAl2009}) and selfing species (\citealt{HusbandAndSchemske1996}) show greater degrees of inbreeding depression than would be expected if deleterious recessive alleles alone were responsible (although numerous very mildly deleterious alleles may explain these findings). Finally, several authors have found evidence of heterozygosity-fitness correlations (HFC's) in wild populations, which are taken as evidence of overdominance (\citealt{LesicaAndAllendorf1992}; \citealt{DaSilvaEtAl2006}; \citealt{FerreiraAndAmos2006}; \citealt{MakinenEtAl2008}; \citealt{HoffmanEtAl2010}).

	One of the appeals of the dominance theory of heterosis is that its mechanisms have firm theoretical foundations. Population genetics theory predicts that recessive deleterious mutations should be common in populations (\citealt{CharlesworthAndWillis2009}). Furthermore, \cite{KacserAndBurns1981} presented theoretical and experimental results that show why deleterious mutations are frequently recessive.

	In contrast, there is no widely accepted mechanism for overdominance and selection for heterozygous genotypes (\citealt{CharlesworthAndWillis2009}; \citealt{ZhouEtAl2012}). This paper attempts to provide such a mechanism, which will provide an explanation for how heterozygosity can result in heterosis even in the absence of deleterious mutations. An interesting feature of the theory presented in this paper is that overdominant loci are inherently epistatic, which may be a reason why papers that find evidence for overdominance also tend to find evidence for epistasis (\citealt{LiEtAl2001}; \citealt{ZhouEtAl2012}).

	The central hypothesis of this paper, that the specificity of protein aggregation reactions provide a physical basis for overdominance and heterozygous advantage, is supported by the findings of previous papers that relate heterozygosity to protein metabolism. The earliest papers, such as \cite{KoehnAndShumway1982} and \cite{HawkinsEtAl1986}, found that both metabolic efficiency and protein turnover increase with decreasing heterozygosity in marine bivalves. More recently, \cite{KristensenEtAl2002} and \cite{PedersenEtAl2005} found that inbred lines of \textit{Drosophila melanogaster} produce higher concentrations of molecular chaperones than outbred lines, which they hypothesized was due to higher rates of protein aggregation. Both \cite{KristensenEtAl2002} and \cite{Goff2011} argued that these previous findings can be explained if homozygosity for deleterious mutations leads to greater expression of unstable proteins by inbred individuals. In contrast, \cite{Ginn2010} attempted to show that the previous findings could be explained by an overdominance model if protein aggregation is assumed to be a highly specific process (see below). \cite{MeadEtAl2003} had already shown that the specificity of prion amyloid formation resulted in balancing selection at the prion protein gene. However, \cite{MeadEtAl2003} and \cite{Ginn2010} only considered the benefits of heterozygosity at a single gene locus. This paper will provide a biochemical explanation for how heterozygosity at multiple gene loci can lead to truncation selection, which can maintain protein polymorphisms at numerous gene loci in natural populations (\citealt{King1967}; \citealt{Milkman1967}; \citealt{SvedEtAl1967}).

	One of the benefits of this paper's theoretical approach is that it can potentially explain several trends in the ploidy level of organisms. The reason why organisms have different ploidy levels is still poorly understood (\citealt{OttoAndGerstein2008}; \citealt{OttoAndWhitton2000}; \citealt{Mable2004}; \citealt{Madlung2013}). Yet, the evolution of different ploidy levels is important to any theory of heterosis since heterozygosity cannot exist in strictly haploid organisms. Most theories that attempt to explain the evolution of higher ploidy levels have focused on the masking of deleterious recessive alleles or have compared the rates of evolution of organisms with different ploidy levels (see \citealt{OrrAndOtto1994}, \citealt{Otto2007}, and \citealt{OttoAndGerstein2008} for review). The theory developed in this paper will instead focus on how higher ploidy levels can help organisms cope with protein aggregation. The theory can potentially explain four different trends: 1) the frequent occurrence of polyploid organisms in harsh environments, 2) the restriction of haploidy to relatively simple organisms, 3) the relative stress tolerances of plant gametophytes and sporophytes, and 4) why complexity (and diploidy) is associated with the production of sexual spores in plants and fungi. Thus, the theory presented in this paper attempts to provide a complete theory of heterozygous advantage that unifies our understanding of heterosis, selection for heterozygosity, and ploidy level.
\clearpage 

\section{Metabolic Heterosis} \label{sec:MetabolicHeterosis}
\subsection{Inbreeding and Metabolic Efficiency} \label{subsec:Efficiency}

	Numerous studies have been published on a phenomena that I will call "metabolic heterosis" in this paper. Metabolic heterosis is the observed correlation between the growth rate of organisms and their heterozygosities as measured by allozyme or microsatellite markers  (\citealt{KoehnAndShumway1982}; \citealt{GartonEtAl1984}; \citealt{Mitton1985}; \citealt{DanzmannEtAl1987}; \citealt{Mitton1993}; \citealt{HedgecockEtAl1996}; \citealt{PogsonAndFevolden1998}; \citealt{BayneEtAl1999}; \citealt{HawkinsAndDay1999}; \citealt{HawkinsEtAl2000}; \citealt{Bayne2004}; \citealt{BorrellEtAl2004}; \citealt{PujolarEtAl2005}; \citealt{KetolaAndKotiaho2009}). Some of these authors maintain that their correlations indicate a relationship between heterozygosity and metabolic efficiency (\citealt{GartonEtAl1984}; \citealt{Mitton1993}; \citealt{Mitton1997}; \citealt{PogsonAndFevolden1998}; \citealt{BorrellEtAl2004}). While most studies have focused on the size and mass of the organism, metabolic efficiency can affect other fitness traits. For instance, \cite{GajardoAndBeardmore1989} and \cite{GajardoEtAl2001} have shown a positive correlation between heterozygosity and the percentage of female \textit{Artemia} that produce energetically expensive encysted offspring rather than energetically cheaper nauplii. While many studies have concluded that heterozygosity is correlated with the metabolic efficiency of organisms, there is still no consensus view on the underlying mechanism behind this correlation. Furthermore, there is an additional debate over whether the fitness parameters are correlated with the heterozygosity of the organism across all gene loci (the ``general effect'' hypothesis) or only with the heterozygosity of gene loci near the measured genetic markers (the ``local effect'' hypothesis) (\citealt{MittonandPierce1980}; \citealt{BallouxEtAl2004}; \citealt{SzulkinEtAl2010}).

	The correlation between heterozygosity and metabolic efficiency may be explained by protein turnover (\citealt{HawkinsEtAl1986}; \citealt{HawkinsEtAl1989}; \citealt{HedgecockEtAl1996}; \citealt{Bayne2004}). \cite{HawkinsEtAl1986} showed that lower heterozygosity leads to higher levels of protein turnover in the blue mussel, \textit{Mytilus edulis}, using $^{15}$N labeled food and allozyme markers. Protein turnover refers to an organism's daily degradation and synthesis of proteins, both of which are energy consuming processes. Therefore, these papers argued, an inbred organism's biomass may be more energetically expensive to sustain than an outbred organism's due to higher levels of protein turnover.

\begin{figure}[!t]
\centering
\includegraphics[scale=0.25]{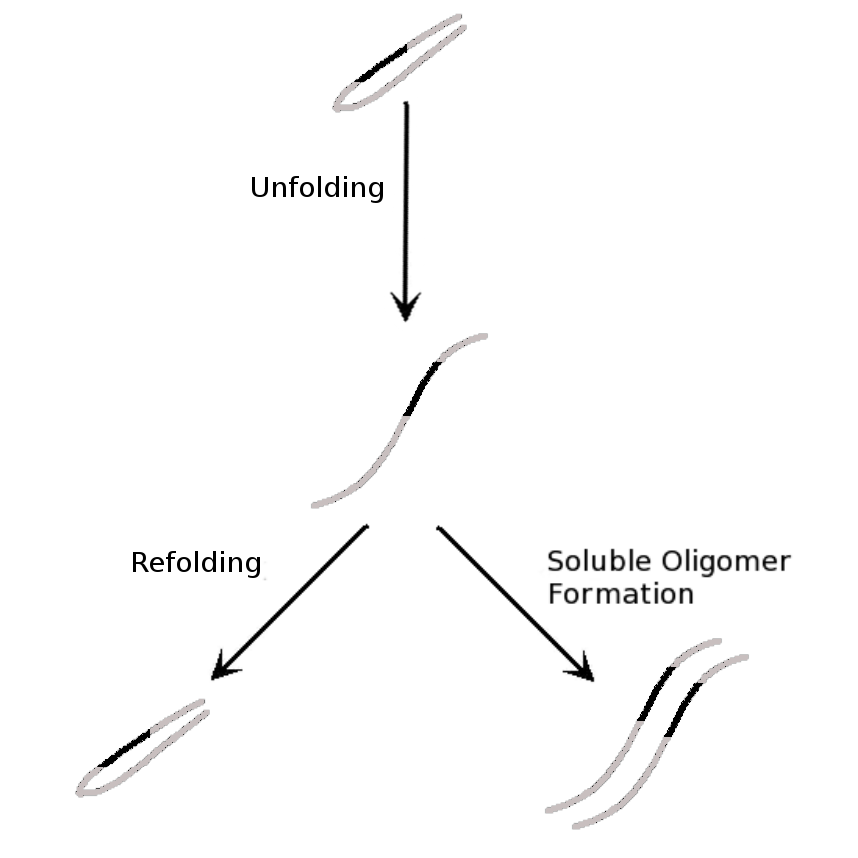}
\caption{An unfolded polypeptide chain may either refold into its native conformation or bind to another unfolded polypeptide chain. The two reactions compete against each other, and the relative rates of each reaction will determine the folding efficiency of the polypeptide chain. The shaded areas on the polypeptide chain represent binding sites that stabilize the folded protein or soluble oligomer.}
\label{fig:Competition}
\end{figure}

	One of the processes that enhances protein turnover is protein aggregation. Most proteins consist of polypeptide chains that must fold into a native conformation in order to be functional. Many proteins, especially meta-stable proteins containing intrinsically unstructured regions (IURs), are continuously unfolding and refolding in an organism (\citealt{OlzschaEtAl2011}; also see Section \ref{subsec:Intrinsic} below). However, as shown in Figure \ref{fig:Competition}, there are two alternative pathways that unfolded polypeptide chains may take. First, the polypeptide chain may refold into its correct conformation and become a functional protein. Alternatively, the polypeptide chain may bind with another unfolded chain and form a soluble oligomer (\citealt{SilowAndOliveberg1997}; \citealt{BitanEtAl2001}; \citealt{Idicula-ThomasAndBalaji2007}; \citealt{KayedEtAl2003}; \citealt{KayedEtAl2004}; \citealt{ClearyEtAl2005}; \citealt{HaassAndSelkoe2007}; \citealt{VieiraEtAl2007}; \citealt{WeiEtAl2007}). Soluble oligomers can then bind with additional unfolded chains and eventually become a solid protein aggregate. The formation of soluble oligomers and solid protein aggregates is detrimental for two reasons. First, protein aggregation competes with the proper folding of a protein (\citealt{KiefhaberEtAl1991}). A protein's folding efficiency decreases when more unfolded polypeptide chains bind to each other. Second, soluble oligomers and solid protein aggregates are cytotoxic species that have been associated with several disorders (\citealt{KayedEtAl2003}; \citealt{HaassAndSelkoe2007}; \citealt{VieiraEtAl2007}; \citealt{ShankarEtAl2008}). Therefore, the viability of organisms depends on the ability of their proteins to maintain their correct conformations and avoid aggregation.

	For this reason, all organisms produce numerous molecular chaperones that prevent unfolded polypeptide chains from aggregating. Molecular chaperones can bind to unfolded polypeptide chains, thereby preventing their aggregation, or they can tag the polypeptide chains with ubiquitin, which marks the polypeptide chains for destruction by the proteasome (\citealt{HayesAndDice1996}; \citealt{Kopito2000}; \citealt{Maurizi2002}; \citealt{McClellenEtAl2005}; \citealt{KaganovichEtAl2008}). The proteasomal system also degrades proteins after they have aggregated (\citealt{DouganEtAl2006}; \citealt{Rubinsztein2006}; \citealt{LiberekEtAl2008}; \citealt{TetzlaffEtAl2008}). In addition, protein aggregates may also be degraded via autophagy, whereby aggregated proteins are transported to lysosomes and digested (\citealt{Kopito2000}; \citealt{Garcia-MataEtAl2002}; \citealt{KruseEtAl2006}; \citealt{YorimitsuAndKlionsky2007}). \cite{ChoeAndStrange2008} observed that half of the genes up-regulated when the nematode \textit{C. elegans} is exposed to aggregate promoting environmental stresses are associated with protein degradation. Especially up-regulated were genes associated with proteasomal and lysosomal degradation.

	Both the molecular chaperones that prevent aggregation and the degradation pathways that destroy soluble oligomers (and protein aggregates) must consume ATP while functioning. In addition, new polypeptide chains will have to be synthesized in order to take the place of degraded chains, so high rates of protein aggregation will also result in greater energy expenditure for protein synthesis. Thus, an organism's efforts to maintain protein homeostasis will generally result in high levels of protein turnover and an associated energy expenditure.

	The expression of molecular chaperones is correlated with the heterozygosity of organisms. \cite{KristensenEtAl2002} and \cite{PedersenEtAl2005}, using an enzyme-linked immunosorbant assay, found that inbred fruit flies (\textit{D. melanogaster} and \textit{Drosophila buzzati}) synthesized more heat shock proteins (\textit{Hsps}) than outbred fruit flies at benign and elevated temperatures. Since \textit{Hsps} are a type of molecular chaperone, the authors of these papers concluded that inbred fruit flies contain a higher number of unfolded or misfolded polypeptide chains, and potentially higher rates of protein aggregation, than outbred fruit flies, even at benign temperatures. Thus, low levels of heterozygosity may result in higher levels of protein aggregation, which in turn can result in higher rates of protein turnover and lower metabolic efficiencies.\footnote{\cite{ChenEtAl2006} has supported these findings with similar experiments performed on Pacific Abalone populations.}

	\cite{KristensenEtAl2002} and \cite{KristensenEtAl2009} used the dominance theory of inbreeding depression to explain the correlation between heterozygosity and lower \textit{Hsp} concentrations. They argued that proteins encoded by deleterious recessive alleles may be less stable, and more prone to aggregation, than the proteins encoded by normal alleles. Consequently, the increased expression of deleterious recessive alleles by inbred organisms may increase their demand for molecular chaperones. The following subsection will develop an overdominance theory to explain the correlations between heterozygosity and \textit{Hsp} concentrations, protein turnover, and metabolic efficiency. The overdominant explanation will then be extended in subsequent sections to provide a biochemical basis for truncation selection that favors heterozygosity. Afterwards, the truncation selection model will be used to explain why higher ploidy levels are advantageous in certain circumstances.

\subsection{Model} \label{subsec:Model}

	A heuristic model is developed in this subsection that shows how an organism's heterozygosity can influence its expression of molecular chaperones, protein turnover, and metabolic efficiency. This model shows how protein aggregation reactions can provide a physical basis for overdominance and metabolic heterosis, which may explain the results of some breeding experiments (\citealt{LiEtAl2001}; \citealt{ZhouEtAl2012}). The model will also show that the relationship between heterozygosity and metabolic efficiency should be linear (additive epistasis) as described in several studies (\citealt{KoehnAndShumway1982}; \citealt{GartonEtAl1984}; \citealt{MittonAndGrant1984}; \citealt{HawkinsEtAl1986}). This contrasts with the usual exponential relationship between heterozygosity and phenotype anticipated by multiplicative epistasis (\citealt{CharlesworthAndWillis2009}).

	The model will assume that an organism maintains steady-state concentrations of functional proteins that are continuously unfolding and either refolding or forming soluble oligomers. Soluble oligomers must be degraded by the organism when they form because they are toxic. Also, the organism must synthesize new proteins to replace those that were removed when the organism destroyed its soluble oligomers. The result is three steady-state concentrations, \textit{[F]$_{steady}$}, \textit{[U]$_{steady}$}, and \textit{[O]$_{steady}$}, which are the concentrations of folded protein, unfolded protein, and soluble oligomer, respectively. In this model, it will be assumed that the organism maintains soluble oligomers at a critical steady-state concentration. If the steady-state concentration of soluble oligomers rises, then the organism will increase its concentration of molecular chaperones to lower the soluble oligomers' steady-state concentration back to their critical level. This may be accomplished through a feedback mechanism, such as the unfolded protein response (UPR) that occurs inside the endoplasmic reticulum (\citealt{SchroderAndKaufman2005}; \citealt{BernalesEtAl2006}). As a consequence of the steady-state assumption, the rates of soluble oligomer formation and degradation will primarily depend on the rate that unfolded polypeptide chains are introduced into the system (see Equation \ref{eq:MaintenanceCalories} below), which is similar to the \textit{in vivo} model described in \cite{KiefhaberEtAl1991}.

	Another assumption for the model is that the initial binding reactions are the rate-limiting step in the various protein aggregation pathways, and that molecular chaperones prevent the accumulation of protein aggregation products ``downstream'' from the initial binding reactions (see \cite{Dobson2003} for an overview of the many aggregation pathways that can occur). This approach has been used by other researchers to successfully model protein aggregation dynamics both \textit{in vitro} and \textit{in vivo} ({\citealt{KiefhaberEtAl1991}; \citealt{HasegawaEtAl1999}; \citealt{BorgiaEtAl2013}). As a consequence of this assumption, the kinetics of soluble oligomer formation will follow second-order rate laws in the model.

	The first thing to consider is the rate at which an unfolded polypeptide chain folds into its native conformation. The polypeptide chain may be unfolded because it has been recently synthesized or because a previously native protein unraveled. The latter process may be part of the protein's normal condition (perhaps because it contains intrinsically unstructured regions) or may be induced by environmental stress. Regardless, most unfolded polypeptide chains must fold into their correct conformation in order to be functional. This takes time, especially if the folding chain becomes trapped in a metastable intermediate state (\citealt{OnuchicEtAl1995}; \citealt{LevyEtAl2005}; \citealt{NevoEtAl2005}).  Nevertheless, folding (and refolding) proceeds according to a first-order rate law (\citealt{KiefhaberEtAl1991}).

\begin{equation}  \label{eq:SimpleFoldRate}
\frac{d[N]}{dt} = k_{f}[U]
\end{equation}
where \textit{[N]} is the concentration of native protein, \textit{t} is time \textit{k$_{f}$} is the rate constant for the folding reaction, and \textit{[U]} is the concentration of unfolded polypeptide chains.

	Alternatively, the unfolded polypeptide chain may bind to another and form a soluble oligomer, which may serve as a seed for protein aggregation. The process of protein aggregation is highly specific in that protein aggregates are highly enriched with a single protein species, even when two or more polypeptides are aggregating simultaneously (\citealt{LondonEtAl1974}; \citealt{SpeedEtAl1996}; \citealt{Kopito2000}; \citealt{RajanEtAl2001}; \citealt{MorellEtAl2008}). The process may be so specific that small differences in amino acid sequence can inhibit co-aggregation of different polypeptide chains. For example, \cite{MeadEtAl2003}, \cite{O'NuallainEtAl2004}, and \cite{ApostolEtAl2010} have found that a single point mutation can prevent amyloid fibrils from co-aggregating. Other researchers have similarly found that amyloid formation can be inhibited in mixtures of polypeptide variants (\citealt{HasegawaEtAl1999}; \citealt{RochetEtAl2000}; \citealt{LashuelEtAl2003}; \citealt{YagiEtAl2005}; \citealt{LewisEtAl2006}; \citealt{Tahiri-AlaouiEtAl2006}).

	Both \cite{O'NuallainEtAl2004} and \cite{ApostolEtAl2010} have argued that the specificity of amyloid formation may be due to changes in amyloid conformation brought about by point mutations. \cite{KingEtAl1996}, \cite{SinhaAndNussinov2001}, and \cite{XuEtAl2013} have shown that point mutations can change the probabilities of unfolded polypeptide chains assuming particular conformations without altering the stability or conformation of the native protein. Thus, point mutations may reduce the probability of two allozymes assuming similar conformations, which would decrease the likelihood of them aligning properly in order to co-aggregate (\citealt{O'NuallainEtAl2004}; \citealt{MaAndNussinov2012}). In addition, point mutations may introduce incompatibilities (e.g. steric repulsion between side chains) that prevent two polypeptide chains from co-aggregating (\citealt{ApostolEtAl2010}), though this would seem to be a less general mechanism. Note that some polymorphisms do co-aggregate (see \cite{WrightEtAl2005} and \cite{KrebsEtAl2004}), but they are not likely to persist in natural populations since they do no confer any heterozygous advantage. Also, the specificity of protein aggregation may only apply to proteins with certain physical properties. This paper will discuss the importance of proteins with intrinsically unstructured regions (IURS) in Sections \ref{subsec:Intrinsic} and \ref{subsec:HaploidDiploid}. However, this is not a major limitation since balancing selection would only apply to a fraction of all protein coding gene loci.

	The specificity of protein aggregation implies that the formation of soluble oligomers proceeds as a second order reaction (\citealt{KiefhaberEtAl1991}; \citealt{BitanEtAl2001}; \citealt{ZhdanovAndKasemo2004}; \citealt{ZhuEtAl2010}). For example, the rate law for the formation of a dimer will be:

\begin{equation}  \label{eq:SimpleOligomerRate}
\frac{d[O]}{dt} = k_{b}[U]^{2}
\end{equation}
where \textit{[O]} is the concentration of the soluble oligomer and \textit{k$_{b}$} is the rate constant for the self-binding reaction.

	A comparison of Equations \ref{eq:SimpleFoldRate} and \ref{eq:SimpleOligomerRate} reveals that the rate of soluble oligomer formation is more dependent upon the concentration of unfolded polypeptide chains than is the rate of protein folding. For example, an individual that is heterozygous at a given gene locus will synthesize two different allozymes, which should have approximately half the concentration that they would have if the individual were homozygous at the gene locus. From Equation \ref{eq:SimpleFoldRate}, it follows that each allozyme will fold half as fast, and the combined rate of folding will be about the same for heterozygous and homozygous individuals. This can be expressed generally as:

\begin{equation}  \label{eq:GeneralFoldRate}
\frac{d[N]}{dt} = \sum{r_{i}k_{fi}[U_{i}]}
\end{equation}
where \textit{r$_{i}$} is the actual concentration of allozyme \textit{i} in an individual divided by the concentration of \textit{i} if the individual were homozygous for \textit{i} (\textit{r$_{i}$} will have a value of 1 in an organism that is homozygous for \textit{i}, a value of 0 in an organism that does not produce \textit{i}, and a value of {$\approx$}0.5 in an organism that is heterozygous for \textit{i}).

	In contrast, soluble oligomers of an allozyme will form at one-quarter the rate in an individual that is heterozygous for the allozyme than in an individual that is homozygous for the allozyme (rate$_{het}\approx$0.25rate$_{hom}$). The combined rate of soluble oligomer formation for allozymes in heterozygous individuals will be approximately half the rate for homozygous individuals (0.25rate$_{hom}$ + 0.25rate$_{hom}$ = 0.5rate$_{hom}$), or more generally:

\begin{equation}  \label{eq:GeneralOligomerRate}
\frac{d[O]}{dt} = \sum{r_{i}^{2}k_{bi}[U_{i}]^{2}}
\end{equation}
Equations \ref{eq:SimpleFoldRate}--\ref{eq:GeneralOligomerRate} show that diluting the concentration of an unfolded polypeptide chain shifts the competition between protein folding and soluble oligomer formation in favor of folding. Thus, heterozygosity increases the folding efficiency of unfolded polypeptide chains simply by diluting their concentrations.

	Another way to think of the influence that heterozygosity has on soluble oligomer formation is to consider the number of collisions that will occur in a given time period. In a homozygous organism, all of the collisions will be between the same allozyme. In contrast, 50\% of the collisions in a heterozygous organism will be between alternate allozymes. Each allozyme buffers the soluble oligomer formation reaction of the other. The critical assumption is that soluble oligomer formation is highly specific, which is corroborated by the research papers cited above.

	Equation \ref{eq:GeneralOligomerRate} gives the rate of dimer formation at a single gene locus. The total rate of soluble oligomer addition for all alleles at all gene loci is:

\begin{dmath}  \label{eq:AdditionRate}
\frac{dA}{dt} = \sum_{j}{\sum_{i}{k_{b_{1}ji}r^{2}_{Uji}[U_{ji}]^{2}}} + \sum_{j}{\sum_{i}{k_{b_{2}ji}r_{Dji}r_{Uji}[D_{ji}][U_{ji}]}} + \sum_{j}{\sum_{i}{k_{b_{3}ji}r_{Tji}r_{Uji}[T_{ji}][U_{ji}]}} \dots
\end{dmath}
where \textit{[U$_{ji}$]} is the concentration of unfolded polypeptide chain expressed by each allele \textit{i} at each gene locus \textit{j}, \textit{[D]} is the concentration of dimer, \textit{[T]} is the concentration of trimer, \textit{r$_{ji}$} is the concentration of a chemical species in an organism that is heterozygous for \textit{ji} divided by its concentration in an organism that is homozygous for \textit{ji}, and \textit{k$_{bji}$} is the rate law constant for the binding reaction of each unfolded polypeptide chain encoded by each allele at each gene locus.

	For the purpose of simplicity, it is assumed in the rest of this subsection that a single molecular chaperone is responsible for removing soluble oligomers. The rate of removal can be expressed in terms of the Michaelis-Menten Equation for the molecular chaperone (\citealt{Kondepudi2008}):

\begin{equation}  \label{eq:RemovalRate}
\frac{dR}{dt} = \frac{R_{max}[O]_{steady}}{K_{m} + [O]_{steady}}
\end{equation}
where \textit{R$_{max}$} is the chaperone's maximum rate of removal, \textit{[O]$_{steady}$} is the steady-state concentration of soluble oligomer, and \textit{K$_{m}$} is the Michaelis-Menten constant for the molecular chaperone. The value of \textit{R$_{max}$} is directly proportional to the concentration of molecular chaperone:

\begin{equation}  \label{eq:Rmax}
R_{max} = k_{2}M_{t}
\end{equation}
where \textit{M$_{t}$} is the total concentration of molecular chaperone and \textit{k$_{2}$} is the rate law constant for the molecular chaperone's catalyzing reaction. Combining Equations \ref{eq:AdditionRate}, \ref{eq:RemovalRate}, and \ref{eq:Rmax} and solving for \textit{M$_{t}$} gives:

\begin{equation}  \label{eq:Chaperone}
M_{t} = (\frac{K_{m} + [O]_{steady}}{k_{2}[O]_{steady}})\frac{dA}{dt}
\end{equation}
Equation \ref{eq:Chaperone} gives the concentration of molecular chaperone necessary to maintain a particular steady-state concentration of soluble oligomers at a given heterozygosity. It predicts that inbred organisms will have higher concentrations of molecular chaperones than outbred organisms, which has been confirmed by \cite{KristensenEtAl2002}.

	The degradation of soluble oligomers must be balanced by the synthesis of new proteins in order to maintain a steady-state concentration of properly functioning protein within the cytosol:

\begin{equation}  \label{eq:SynthesisRate}
\frac{dP}{dt} = \nu\frac{dR}{dt} = \nu\frac{dA}{dt}
\end{equation}
where \textit{dP/dt} is the rate that new proteins are synthesized to replace proteins that have formed soluble oligomers and \textit{$\nu$} is the number of proteins making up the soluble oligomers. Equations \ref{eq:AdditionRate}, \ref{eq:RemovalRate}, and \ref{eq:SynthesisRate} predict that inbred organisms should have higher rates of protein turnover than outbred organisms, which has been confirmed by \cite{HawkinsEtAl1986}.

	Finally, the removal of soluble oligomers and their replacement with new proteins is an energy consuming process. The overall calorie consumption rate due to protein maintenance, \textit{dC$_{Maint}$/dt}, is:

\begin{dmath}  \label{eq:MaintenanceCalories}
\frac{dC_{Maint}}{dt} = \sum_{j}{\sum_{i}{(\beta_{ji} + 2\gamma_{ji})k_{b_{1}ji}r^{2}_{Uji}[U_{ji}]^{2}}} + \sum_{j}{\sum_{i}{(\beta_{ji} + 3\gamma_{ji})k_{b_{2}ji}r_{Dji}r_{Uji}[D_{ji}][U_{ji}]}} + \sum_{j}{\sum_{i}{(\beta_{ji} + 4\gamma_{ji})k_{b_{3}ji}r_{Tji}r_{Uji}[T_{ji}][U_{ji}]}} \dots
\end{dmath}
where \textit{$\beta$} is the number of calories consumed during the degradation of each soluble oligomer and \textit{$\nu\gamma$} is the number of calories consumed during the synthesis of each replacement protein. Equation \ref{eq:MaintenanceCalories} predicts that inbred organisms should be less metabolically efficient than outbred organisms, which has been confirmed by numerous studies (see \citealt{Mitton1997} for review). A simplified version of Equation \ref{eq:MaintenanceCalories} is obtained by assuming that only the dimerization reaction has a non-negligible rate, and by assigning every aggregating protein the same value for \textit{k$_{b}$} and \textit{[U]$^{2}$}, which are then combined into a single parameter, \textit{k$_{1}$}:

\begin{equation} \label{eq:SimpleMaintenanceCalories}
\frac{dC_{Maint}}{dt} = k_{1}[\beta + 2\gamma][T - 0.5N_{Het}]
\end{equation}
where T is the total number of gene loci that synthesize aggregation-prone proteins and N$_{Het}$ is the number of heterozygous gene loci. As shown in Figure \ref{fig:MaintenanceCalories}, Equation \ref{eq:SimpleMaintenanceCalories} is linear with respect to heterozygosity, which means that the overdominant loci exhibit modest epistasis for metabolic heterosis, otherwise the equation should decline exponentially with heterozygosity (\citealt{CharlesworthAndWillis2009}). A linear relationship between metabolic efficiency (as measured by oxygen consumption and growth rate) and heterozygosity has been found for several species of marine bivalves (\citealt{KoehnAndShumway1982}; \citealt{GartonEtAl1984}; \citealt{MittonAndGrant1984}; \citealt{HawkinsEtAl1986}).

\begin{figure}[!t]
\centering
\includegraphics[scale=0.29]{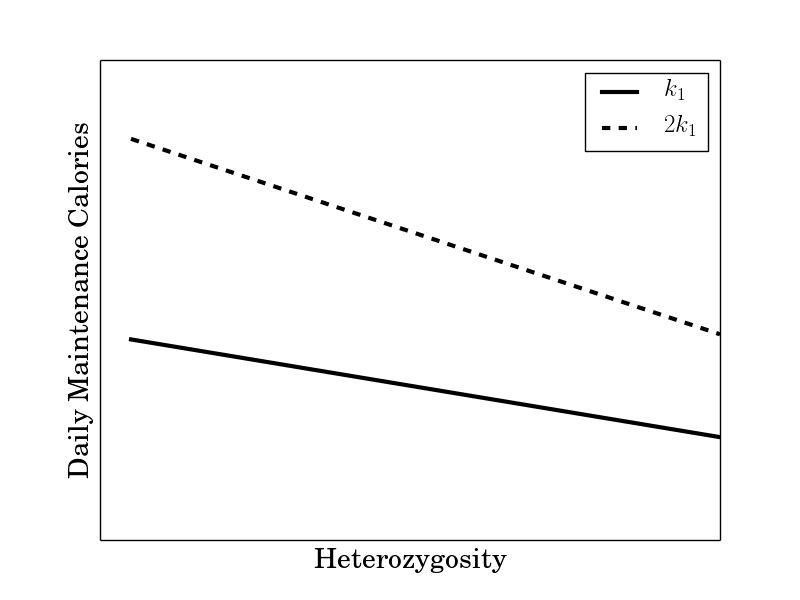}
\caption{Results from Equation \ref{eq:SimpleMaintenanceCalories} showing that metabolic efficiency increases with heterozygosity. The dashed curve shows the results for a higher \textit{k$_{1}$} value, which increases with both the stressfulness of the environment and the abundance of aggregation-prone proteins synthesized by the organism.}
\label{fig:MaintenanceCalories}
\end{figure}

	Equations \ref{eq:SimpleFoldRate}--\ref{eq:MaintenanceCalories} provide a heuristic model for the relationship between protein aggregation and metabolic heterosis. The model shows that heterozygosity can be beneficial even in the absence of deleterious mutations, and is in agreement with previous research that found a correlation between heterozygosity and: (1) metabolic efficiency (\citealt{KoehnAndShumway1982}; \citealt{GartonEtAl1984}; \citealt{Mitton1997}), (2) protein turnover (\citealt{HawkinsEtAl1986}; \citealt{HawkinsEtAl1989}), and (3) expression of molecular chaperones (\citealt{KristensenEtAl2002}; \citealt{PedersenEtAl2005}). As shown in Figure \ref{fig:MaintenanceCalories}, the model predicts that inbreeding depression should increase with the stressfulness of the environment because the values of \textit{k$_{b}$} and \textit{[U]$_{steady}$} for each allele should increase with the stressfulness of the environment (see Subsection \ref{subsec:Stress}). This is also supported by previous research (see \citealt{ArmbrusterAndReed2005} for review).

\subsection{Thermodynamic Considerations} \label{subsec:Thermodynamics}

	It may be helpful to conceptualize the underlying physics of the model developed in this paper. The model assumes steady-state conditions in which the rates of soluble oligomer addition, soluble oligomer removal, and synthesis of replacement proteins are equal (Equation \ref{eq:SynthesisRate}). This results in steady-state concentrations of unfolded polypeptide chains, soluble oligomers, and native proteins. These steady-state concentrations are not in chemical equilibrium, otherwise the reactions for soluble oligomer formation would not proceed forward. For example, the fact that two unfolded polypeptide chains bind to each other to form dimers suggests that they have greater chemical potentials than the dimers (i.e. $2\mu_{unfolded} > \mu_{dimer}$). We can express this in terms of chemical affinity, \textit{A}, which is the difference in chemical potential between reactants and products (\citealt{Kondepudi2008}).

\begin{equation} \label{eq:affinity}
A \equiv 2\mu_{unfolded} - \mu_{dimer} > 0
\end{equation}
The greater the chemical affinity of a reaction, the further away it is from equilibrium.

	However, the aggregation reactions proceed forward more slowly in hybrid organisms than in non-hybrids because the hybrids produce more allozymes. Point mutations can change the probability distributions for unfolded polypeptide chains assuming particular conformations without affecting the stability of the polypeptides' native conformations (\citealt{KingEtAl1996}; \citealt{SinhaAndNussinov2001}; \citealt{XuEtAl2013}). This lowers the probability of two allozymes co-aggregating because polypeptide chains must have similar conformations in order to form the necessary cross-$\beta$ interactions that allow them to bind to each other (\citealt{O'NuallainEtAl2004}). Point mutations in polypeptide chains can also introduce incompatibilities (e.g. steric repulsion between side chains) that prevent two chains from co-aggregating (\citealt{ApostolEtAl2010}).  Therefore, hybrid organisms benefit from lower rates of protein aggregation due to their greater degree of ``mixedupness.''

        This can be thought of in terms of chemical potential and chemical affinity. The chemical potentials of the unfolded chains and dimers are related to to their concentrations by:

\begin{subequations} \label{eq:potential}
\begin{align}
\mu_{unfolded} &= \mu^{\circ}_{unfolded} + kTln[U] \\
\mu_{dimer} &= \mu^{\circ}_{dimer} + kTln[D]
\end{align}
\end{subequations}
where \textit{$\mu^{\circ}$} is the standard-state chemical potential, \textit{kT} is the temperature, \textit{$[U]$} is the concentration of unfolded polypeptide chain, and \textit{$[D]$} is the concentration of dimer. Since an organism's molecular chaperones work to keep the concentration of dimer low, and the concentration of unfolded polypeptide chain should be about half as much in an organism that is heterozygous for the polypeptide than in an organism that is homozygous for the polypeptide, then the chemical affinity of the dimerization reaction should be lower in the heterozygous organism than in the homozygous organism. In other words, the homozygous organism is further away from chemical equilibrium.

Chemical affinity can be related to the rates of elementary step reactions by (\citealt{Kondepudi2008}):

\begin{equation} \label{eq:affinityrate}
\frac{A}{kT} = ln(\frac{R_{f}}{R_{r}})
\end{equation}
where \textit{$R_{f}$} is the rate of the forward reaction and \textit{$R_{r}$} is the rate of the reverse reaction. Finally, the relationship between the rate of progression and affinity for a chemical reaction is:

\begin{equation} \label{eq:affinityprogress}
\frac{d\xi}{dt} = R_{f}(1 - e^{\frac{-A}{kT}})
\end{equation}
where the rate of chemical progression, \textit{$\frac{d\xi}{dt}$}, is the difference between the forward and reverse reactions. Comparing Equations \ref{eq:affinity}, \ref{eq:potential}, \ref{eq:affinityrate}, and \ref{eq:affinityprogress}, it follows that the rate of progression for a polypeptide chain's dimerization reaction will be slower in an organism that is heterozygous for the polypeptide than in an organism that is homozygous for the polypeptide because the heterozygous organism is closer to chemical equilibrium.

	Thus, organisms must expend energy to remove soluble oligomers because they maintain steady-state concentrations of native proteins, unfolded polypeptide chains, and soluble oligomers that are different from their equilibrium values. However, the aggregation reactions proceed toward chemical equilibrium more slowly in hybrid organisms than in non-hybrids, so the hybrids need a lower rate of calorie consumption to maintain steady-state conditions (Equation \ref{eq:MaintenanceCalories}).

	An analogy can be drawn with refrigerators, which consume energy to maintain steady-state thermal gradients. The amount of power a refrigerator consumes increases with the difference between its inside and outside temperatures because the rate of inward heat flow increases with the refrigerator's thermal gradient. Likewise, the constant movements toward and away from chemical equilibrium are responsible for organisms' maintenance costs, and the different rates of these movements give rise to the differences in performance (growth rate, size, etc.) between hybrid and non-hybrid organisms.

\section{Epistasis and Truncation Selection} \label{sec:TruncationSelection}

\subsection{Inbreeding and Epistasis} \label{subsec:Epistasis}

\begin{figure}
\centering
\subfigure[]{\includegraphics[scale=0.29]{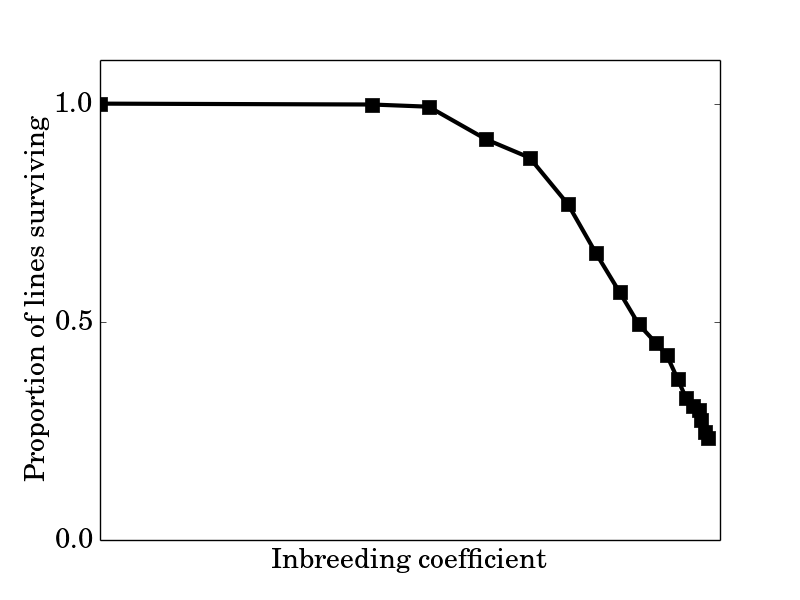}}
\subfigure[]{\includegraphics[scale=0.29]{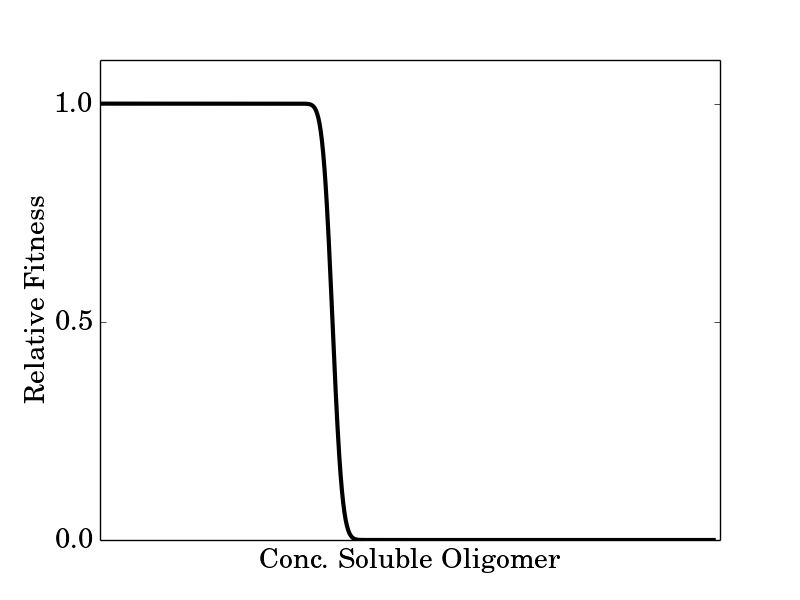}}
\caption{(a) Reproduction of data in Fig. 2 of \cite{RumballEtAl1994}. The curve shows the number of offspring produced by \textit{D. melanogaster} after multiple generations of full-sib mating. (b) Equation \ref{eq:TruncationSelection} gives the relative fitness of an organism with increasing concentration of soluble oligomers.}
\label{fig:Epistasis}
\end{figure}

\begin{figure}
\centering
\includegraphics[scale=0.29]{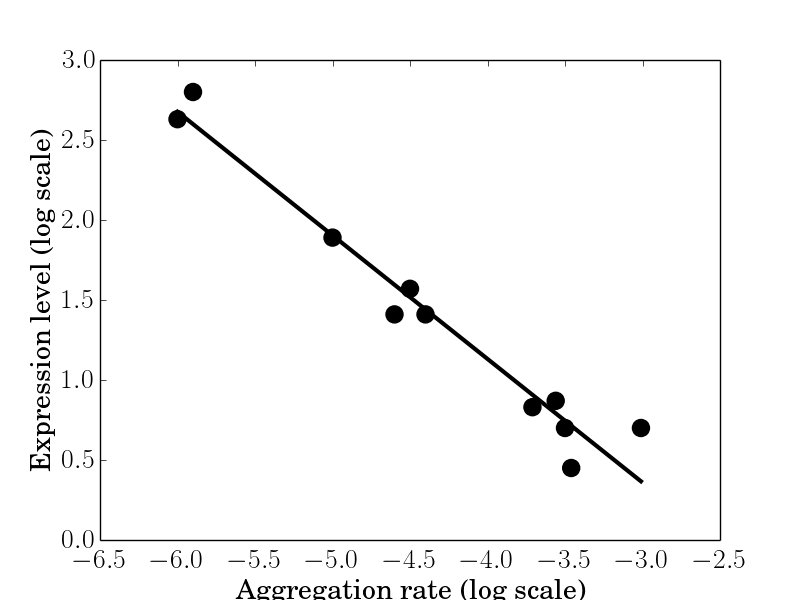}
\caption{Reproduction of plot in Fig. 1 of \cite{TartagliaEtAl2007}. The curve shows that the expression level of a protein is negatively correlated with its aggregation rate. This may mean that organisms produce as much protein as they can before it starts to aggregate.}
\label{fig:Expression}
\end{figure}

	Subsection \ref{subsec:Model} discussed how heterozygosity affects an organism's metabolic performance when the organism maintains steady-state conditions. This subsection will consider the effect heterozygosity has on an organism's viability when the organism's defenses against protein aggregation are overwhelmed and steady-state conditions are no longer maintained. Some studies have found a steep drop in viability at high levels of inbreeding depression, especially in \textit{D. melanogaster} (\citealt{Kosuda1972}; \citealt{RumballEtAl1994}; Figure \ref{fig:Epistasis}). These results suggest that extinction of an inbreeding line is likely to occur when the level of inbreeding exceeds a threshold value (\citealt{Frankham1995}). This might be evidence for synergistic epistasis between deleterious mutations (\citealt{Kosuda1972}; \citealt{Charlesworth1998}). However, it could also be evidence for epistasis between heterozygous gene loci, which \cite{CheloAndTeotonio2012} have observed in their experiments. Several authors have shown that molecular chaperones can provide a biochemical basis for epistasis by inhibiting protein aggregation (\citealt{FaresEtAl2002}; \citealt{SollarsEtAl2003}; \citealt{Maisnier-PatinEtAl2005}; \citealt{DeVisserAndElena2007}; \citealt{DeVisserEtAl2011}; \citealt{Lehner2011}). This subsection expands on these ideas and develops a theory that provides a biochemical basis for the steep decline in viability that sometimes occurs with increasing inbreeding. Then it discusses the model's implications for selection for heterozygosity.

        One of the reasons that protein aggregation is such a nuisance is that organisms seem to produce as much of a protein as they can just before it starts to aggregate. \cite{TartagliaEtAl2007} provides evidence for this in a plot similar to that shown in Figure \ref{fig:Expression}, which shows that the logarithm of a protein's aggregation rate is negatively correlated with the logarithm of its expression level. From this plot, the authors of \cite{TartagliaEtAl2007} conclude that organisms have ``no scope for dealing with any situation in which these [expression] levels rise further or whereby the aggregation rates are increased... In the context of protein solubility, therefore, we are constantly living our lives at the edge of a molecular precipice.'' In other words, we should expect steep declines in fitness when protein homeostasis is perturbed and soluble oligomers accumulate within an organism.

        Since soluble oligomers are toxic substances (\citealt{KayedEtAl2003}; \citealt{HaassAndSelkoe2007}; \citealt{VieiraEtAl2007}; \citealt{ShankarEtAl2008}), the equations used by toxicologists may be useful in modeling the fitness declines brought about by the accumulation of soluble oligomers. The responses of organisms to toxic substances typically follow log-normal distributions (\citealt{WagnerAndLokke1991}; \citealt{WheelerEtAl2002}; \citealt{NewmanAndUnger2003}). Thus, the fitness of individuals with a given concentration of soluble oligomer can be obtained in manner similar to that given in \cite{KimuraAndCrow1978}:

\begin{equation}  \label{eq:TruncationSelection}
W([O]) = W_{max}\sqrt{\frac{1}{2\pi\sigma^{2}}}\int_{[O]}^{\infty}{\frac{1}{C}e^{\frac{-(lnC - lnC_{50})^{2}}{2\sigma^{2}}}dC}
\end{equation}
where \textit{W([O])} is the fitness of individuals with total soluble oligomer concentration \textit{[O]}, \textit{W$_{max}$} is the fitness of individuals that produce no soluble oligomers, \textit{C} is the concentration of soluble oligomer, \textit{C$_{50}$} is the concentration of soluble oligomer that kills half of the individuals in a population, and $\sigma$ is the shape parameter for the log-normal distribution.

	I used a log-normal distribution for Equation \ref{eq:TruncationSelection} because log-normal distributions are commonly used in the toxicology literature, but any distribution that yields an S-shaped CCDF (e.g. normal, log-logistic, Weibull) can also be used in a truncation selection model. Use of a log-normal distribution would be justified if the toxicities of soluble oligomers are due to many random variables that result in multiplicative degradation (\citealt{NISTSEMATECH2003}), which would be true if, as hypothesized by others, their toxic effects are due to a general disruption of homeostasis caused by the permeabilization of cellular and organelle membranes (\citealt{KayedEtAl2003}; \citealt{BucciantiniEtAl2004}; \citealt{KayedEtAl2004}; \citealt{Glabe2006}). A log-normal distribution would also be justified if the aggregation of regulatory and signaling proteins leads to a disruption in developmental homeostasis as postulated by I. Michael Lerner (see Section \ref{subsec:HaploidDiploid}.

	Equation \ref{eq:TruncationSelection} assumes that neither allele at a heterozygous gene locus negatively impacts the phenotype of its carrier. This may not be true and may be addressed with:

\begin{equation}  \label{eq:Suboptimal}
W([O]) = S([O])\prod{(1 - h_{i}s_{i})}\prod{(1 - s_{i})}
\end{equation}
where S([O]) is the right side of Equation \ref{eq:TruncationSelection}, \textit{s} is the fitness cost of the inferior alleles, and \textit{h} is the dominance of the inferior alleles (\textit{h}=1 is completely dominant and \textit{h}=0 is completely recessive). \textit{$\Pi$(1-\textit{hs}}) is the fitness cost of inferior alleles at heterozygous gene loci, and \textit{$\Pi$(1-\textit{s}}) is the fitness cost of inferior alleles at homozygous gene loci.

	The relative fitness of an individual, (\textit{W[O]/W$_{max}$}), can be calculated once the total concentration of soluble oligomers in the individual is known (Figure \ref{fig:Epistasis}). There are two approaches to calculate the concentration of soluble oligomers for Equation \ref{eq:TruncationSelection}. The first approach assumes steady-state conditions like in the previous section, but imposes an upper limit to either the amount of molecular chaperone an organism can produce or to the amount of energy available for the chaperone to perform its function\footnote{ Some chaperones, such as \textit{Hsp70}, require ATP to bind to unfolded polypeptide substrates (\citealt{PattersonAndHohfeld2006}; \citealt{LotzEtAl2010})}. The assumption of an upper limit on available energy is reasonable because there are limits to the amount of oxygen organisms can acquire from their environment (\citealt{Portner2001}; \citealt{BicklerAndBuck2007}; \citealt{RamirezEtAl2007}). This assumption may apply to organisms in moderately severe environments or to severely inbred animals (e.g. \citealt{RumballEtAl1994}). Under such conditions, the rate of soluble oligomer removal (Equation \ref{eq:RemovalRate}) may still equal the rate of soluble oligomer addition (Equation \ref{eq:AdditionRate}), but the effective concentration of molecular chaperone (the concentration of molecular chaperone activated by ATP) is constant, which results in a maximum rate of soluble oligomer removal, (\textit{R$_{max}$}). The result is:

\begin{equation}  \label{eq:MichaelisMenten}
[O]_{steady} = \frac{(K_{m})(\frac{dA}{dt})}{R_{max} - \frac{dA}{dt}}
\end{equation}
The steady-state concentration of soluble oligomer increases as \textit{dA/dt} approaches the value of \textit{R$_{max}$}. This will occur as the stressfulness of the environment increases or with decreasing heterozygosity.

	The second approach assumes that steady-state conditions are disrupted, which would occur if \textit{$\frac{dA}{dt}$} $>$ \textit{R$_{max}$}, or if an organism is unable to remove soluble oligomers once they have formed. There are several reasons why an organism would fail to maintain steady-state conditions:
\begin{enumerate}
\item The organism is dormant. Many organisms possess dormant resting stages, such as spores and cysts, that offer protection during adverse environmental conditions. These resting stages are usually associated with high concentrations of compatible solutes (e.g. trehalose) and small heat-shock proteins (sHSPs). Both compatible solutes and sHSPs can prevent unfolded polypeptide chains from binding to each other without consuming ATP. However, they cannot re-fold, disaggregate, or degrade misfolded proteins; they simply inhibit the formation of soluble oligomers (\citealt{SingerAndLindquist1998}; \citealt{Garay-ArroyoEtAl2000}; \citealt{WatersEtAl2008}; \citealt{VanLeeuwenEtAl2013}).

\item The organism does not have enough readily available energy to maintain protein homeostasis. All organisms have a finite energy supply available to them at any time. The ability of animals to generate ATP, for example, is limited by oxygen availability. Furthermore, several stresses are known to reduce an organism's ability to generate ATP. Thermal stresses can reduce the aerobic scope of ectothermic animals and desiccation can lead to a suspension of metabolism (\citealt{Portner2001}; \citealt{AlpertAndOliver2002}).

\item The organism is exposed to a sudden environmental shock. This can lead to elevated levels of protein aggregation. The organism should respond to the shock by synthesizing more molecular chaperones, but soluble oligomers can accumulate while the organism is adjusting to the new conditions.

\item The organism has an extracellular space. Multicellular organisms can secrete proteins into their extracellular space, which has orders of magnitude lower ATP than the intracellular space. Little is currently known about the process of preserving protein homeostasis in the extracellular space, but animals appear to produce extracellular chaperones that inhibit formation of soluble oligomers in a manner similar to sHSPs (\citealt{PoonEtAl2002}; \citealt{ManniniEtAl2012}; \citealt{WyattEtAl2013}).
\end{enumerate}
In this case, the net rate of soluble oligomer accumulation is given by:

\begin{equation}  \label{eq:Variable}
(\frac{dA}{dt})_{Net} = \frac{dA}{dt} - \frac{dR}{dt}
\end{equation}
Integrating Equation \ref{eq:Variable} will give the concentration of soluble oligomer at a given time, but that requires knowledge of how \textit{[U]} and \textit{k$_{b}$} change with time for each polypeptide chain. The value of \textit{dR/dt} will also change with time if the organisms respond to the environmental shock by increasing the concentrations of their molecular chaperones. Nevertheless, a \textit{[O]} value can be used in Equation \ref{eq:TruncationSelection} once it has been obtained. This would give the survivorship of individuals with different heterozygosities after exposure to an environmental shock for a given length of time. Equation \ref{eq:Variable} does not yield simple solutions like the steady-state equations, but exposures to suddenly elevated stresses may be a more frequent source of viability declines in natural environments.

	Since Equation \ref{eq:AdditionRate} predicts that the rate of soluble oligomer addition increases with decreasing heterozygosity, the above model predicts that viability should decrease with [decreasing]\footnote{the published manuscript contains a word omission here that escaped peer review. I have added the word ``decreasing'' to fix the error.} heterozygosity under a given set of environmental conditions. In general, the steep decline in viability should occur at higher heterozygosities with increasing environmental harshness. This may explain why hybrid crops are more drought tolerant than non-hybrids (\citealt{Duvick2001}), and why a threshold survivorship is seen in some inbreeding studies (\citealt{RumballEtAl1994}; \citealt{Frankham1995}). The model also has relevance for explaining truncation selection for heterozygosity.

\subsection{Truncation Selection for Heterozygosity} \label{subsec:TruncationSelection}

	\cite{LewontinAndHubby1966} provided an influential argument against the hypothesis that natural selection favors heterozygous genotypes. They measured the amount of allozyme diversity in wild \textit{Drosophila pseudoobscura} populations and found polymorphisms segregating at approximately one-third ($\approx$ 2000) of \textit{D. pseudoobscura}'s gene loci. \cite{LewontinAndHubby1966} argued that such a large number of polymorphisms could not be maintained by balancing selection without enormous fitness costs. For example, if homozygosity at a single gene locus reduces the reproductive potential of an individual by 10\%, and only two polymorphisms are segregating at the gene locus at a frequency of 50\% each, then the reproductive potential of the whole population will be reduced by 5\% for each gene locus. The reproductive fitness of the population would be 0.95$^{2000}$, or 10$^{-46}$, its maximum value. This is an unrealistically low number, and they concluded that natural selection could not favor heterozygotes.

\begin{figure}[!t]
\centering
\subfigure[]{\includegraphics[scale=0.29]{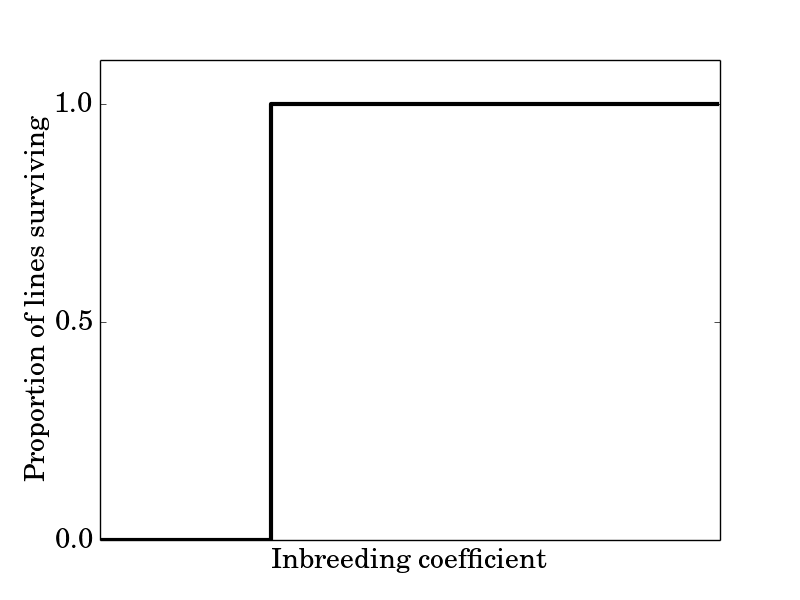}}
\subfigure[]{\includegraphics[scale=0.29]{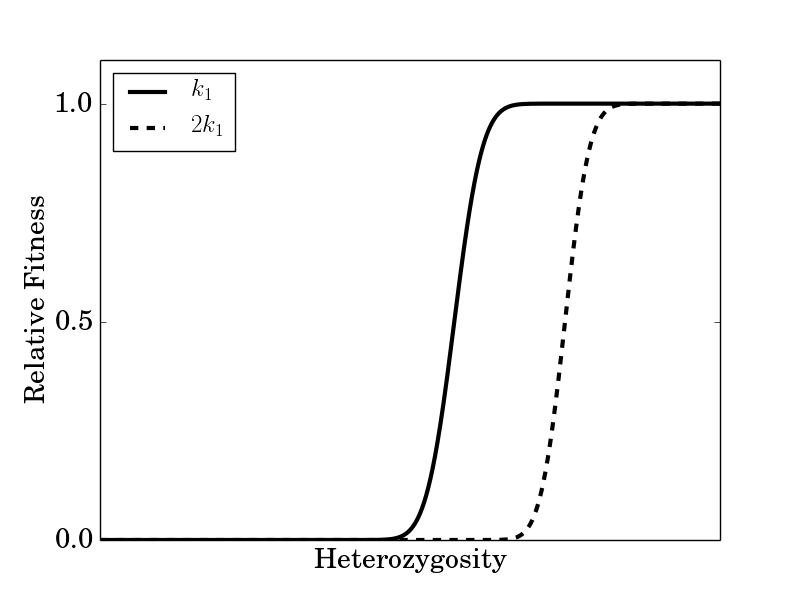}}
\subfigure[]{\includegraphics[scale=0.29]{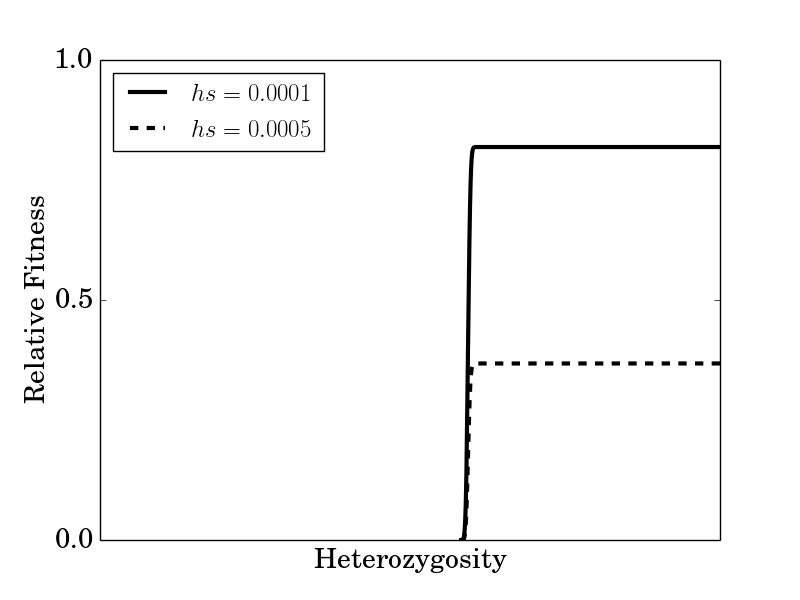}}
\caption{(a) Idealized truncation selection curve. The relative fitness of an organism is 1 above a critical heterozygosity and 0 below the critical heterozygosity. (b) Truncation selection for heterozygosity generated by combining Equations \ref{eq:TruncationSelection} and \ref{eq:MichaelisMenten}. The two curves show truncation selection for different levels of environmental stress as quantified by a \textit{k$_1$} parameter (see main text).(c) Equation \ref{eq:Suboptimal} incorporates the possibility that one of the alleles at a gene locus negatively impacts the phenotype of its carrier.}
\label{fig:TruncationSelection}
\end{figure}

	Shortly after \cite{LewontinAndHubby1966} was published, three papers responded with a similar solution to the problem it raised (\citealt{King1967}; \citealt{Milkman1967}; \citealt{SvedEtAl1967}; \citealt{Crow1992}). These papers proposed that truncation selection could maintain a large number of polymorphisms in natural populations without unreasonable fitness costs. In their models, all individuals whose heterozygosities are below a critical value have a fitness of zero, and individuals whose heterozygosities are above the critical value have maximum fitness (Figure \ref{fig:TruncationSelection}a). \cite{Wills1978} expanded on these models and showed that the number of polymorphisms that can be maintained in a population by truncation selection depends on the effective population size. However, he found that truncation of the least heterozygous individuals (the bottom 5\%) in a population can maintain polymorphisms at 66,000 gene loci in a population of 100,000 individuals, which is more than enough to support the number of polymorphisms that actually occur in natural populations. The effectiveness of truncation selection comes from its severity and its ability to operate on many gene loci simultaneously. Several studies have found evidence for truncation selection acting on natural populations, but the results have been mixed (see \citealt{Mitton1997} for review and \citealt{KaeufferEtAl2007} as a recent example).

	Truncation selection can be considered an extreme form of epistasis in which heterozygosity confers decreasingly small fitness gains with each additional heterozygous gene locus. Both \cite{RumballEtAl1994} and \cite{CheloAndTeotonio2012} have provided experimental evidence that such epistasis exists. The epistasis can be modeled using a S-shaped truncation curve obtained by combining Equations \ref{eq:TruncationSelection} and \ref{eq:AdditionRate} with either \ref{eq:MichaelisMenten} or \ref{eq:Variable} (Figure \ref{fig:TruncationSelection}b). The curve is not strictly a truncation selection curve, but \cite{KimuraAndCrow1978} has shown that selection is almost as efficient when it follows a S-shaped cumulative distribution curve (CDF or CCDF). In order to generate the figure, I assigned the same values of \textit{k$_b$} and \textit{[U]$^2$} to all of the polypeptide chains in Equation \ref{eq:AdditionRate} (I assumed the formation of trimers and tetramers was negligible). These can be considered the average values of \textit{k$_b$} and \textit{[U]$^2$} for aggregation-prone proteins in a hypothetical organism. These values were combined into a single parameter \textit{k$_1$=k$_b$[U]$^2$}, which quantifies the propensity of proteins to aggregate. The value of \textit{k$_1$} should increase with the stressfulness of the environment and the susceptibility of an organism's proteins to aggregation (see below). Figure \ref{fig:TruncationSelection}c shows the results when Equation \ref{eq:Suboptimal} is used to generate a truncation curve. The same value of \textit{hs} is assigned to each suboptimal allele, which can be taken as an average of the \textit{hs} values for all suboptimal alleles.

	The model depicted in Figures \ref{fig:TruncationSelection}b and \ref{fig:TruncationSelection}c may provide a physical basis for truncation selection for heterozygosity. According to the model, organisms are periodically exposed to stresses that lead to an accumulation of soluble oligomers, which in turn results in a sharp decline in viability. Only highly heterozygous individuals will be found in stressful environments because less heterozygous individuals will be on the wrong side of the truncation curve. This illustrates the severity of truncation selection, which is not typically associated with gradual evolution. If all the individuals in a species are on the wrong side of a truncation curve, then the species will go extinct. Thus, some individuals must be on the right side of the truncation curve prior to the species being subjected to truncation selection. In other words, some of the individuals in the species must be ``pre-adapted'' to the environment.

        Another interesting property of the truncation selection model is that it depends only on the number of heterozygous gene loci, and not on the identity of the gene loci. This is analogous to a ``colligative property'' in chemistry. This feature is important because it makes selection for heterozygosity compatible with sexual reproduction. Offspring are not going to be heterozygous at the same gene loci as their parents, but they will have, on average, the same number of heterozygous gene loci as their parents (assuming random mating in the population), so they should have the same overall fitness. Truncation selection allows organisms to substitute heterozygosity at one gene locus for heterozygosity at another gene locus without suffering significant fitness costs because the selection is acting on the overall level of heterozygosity, not on heterozygosity at any particular gene locus. The colligative nature of heterozygous advantage also applies to Equation \ref{eq:SimpleMaintenanceCalories}, which relates heterozygosity to the metabolic efficiency of an organism. Again, overall metabolic efficiency depends on the number of heterozygous gene loci, not on the identity of the gene loci. Of course, there is the caveat that only some of the proteins produced by an organism are aggregation-prone, thus the identity of the gene loci is not completely unimportant. Rather, the colligative nature of heterozygous advantage is confined to the subset of gene loci that code for aggregation-prone proteins.

	The truncation selection model has one important consequence. Factors that increase the accumulation of soluble oligomers in organisms should lead to truncation selection for higher heterozygosities. These factors can be broken up into two categories: 1) environmental stresses that promote protein aggregation, and 2) inherent characteristics of proteins that affect their propensity to aggregate. Section \ref{sec:Ploidy} will discuss how these factors interact to create selection for different ploidy levels, so it will be useful to describe how each promotes protein aggregation.

\subsection{Environmental Stress} \label{subsec:Stress}

	Temperature, water stress, and hypoxia can promote protein aggregation. High and low temperatures denature proteins, which leads to high concentrations of unfolded polypeptide chains (\citealt{BecktelAndSchellman1987}). The values of the rate law constants for protein aggregation also increase with temperature (\citealt{WangAndRoberts2013}). Water stress (desiccation, freezing, salinity\footnote{Both high and low salinities can cause protein aggregation (e.g. \citealt{Chang2005}; \citealt{DownsEtAl2009}; \citealt{TineEtAl2010}; \citealt{MonariEtAl2011})}) is created by the limitation of liquid water inside the organism, which results in high concentrations of unfolded polypeptide chain (because the volume of solvent is limited) (\citealt{GoyalEtAl2005}). Water stress can also cause proteins to unfold by weakening the hydrophobic effect (\citealt{PrestrelskiEtAl1993}; \citealt{AllisonEtAl1999}). In addition, water stress results in molecular crowding, which increases the value of the rate law constants for protein aggregation (\citealt{Ellis2001}; \citealt{SmallwoodAndBowles2002}; \citealt{ChebotarevaEtAl2004}; \citealt{EllisAndMinton2006}; \citealt{WhiteEtAl2010}). Combining the effects of temperature and water stress is particularly harsh. For example, cold temperatures can cause proteins to denature, which greatly enhances the rate of protein aggregation when combined with freezing (\citealt{FranksEtAl1990}; \citealt{SmallwoodAndBowles2002}; \citealt{DiasEtAl2009}; \citealt{SinghEtAl2009}). Likewise, combining high temperatures with arid conditions should also promote aggregation. Finally, hypoxia can hinder the ability of organisms to remove soluble oligomers as they form because many molecular chaperones require ATP to bind to unfolded polypeptide chains (\citealt{PattersonAndHohfeld2006}; \citealt{LotzEtAl2010}). The physical stresses that promote protein aggregation should cause the truncation selection curves depicted in Figure \ref{fig:TruncationSelection} to shift to higher heterozygosity values, so only highly heterozygous individuals should be able to live in harsh environments.

\subsection{Protein Length and IURs} \label{subsec:Intrinsic}

	The intrinsic characteristics of proteins can also affect their propensity to aggregate. \cite{OlzschaEtAl2011} determined what structural features increase a protein's susceptibility to aggregation using molecular templates that seed protein aggregation in human cells. They found that protein size and the presence of intrinsically unstructured regions (IURs) were the two characteristics that most enhanced a protein's susceptibility to aggregation. Both of these characteristics are associated with proteins that contain multiple domains (\citealt{DunkerEtAl2005}; \citealt{FongAndPanchenko2010}).

	A domain is a sequence of amino acids that, if separated from the rest of the polypeptide chain, would still fold into its proper conformation and function normally. A multi-domain protein is a protein that contains multiple domains. Some multi-domain proteins can be considered a string of proteins that are joined together. In fact, proteins that exist separately in some species may be found as parts of multi-domain proteins in other species, a phenomena called ``domain accretion" (\citealt{KooninEtAl2002}; \citealt{BasuEtAl2009}). Multi-domain proteins are important because their modularity makes them capable of adapting to rapidly changing environments (\citealt{SunAndDeem2007}; \citealt{HeEtAl2009}; \citealt{LorenzEtAl2011}; \citealt{ParkEtAl2015}). As expected, multi-domain proteins tend to be larger than single-domain proteins (\citealt{TanEtAl2005}; \citealt{TordaiEtAl2005}).

	Multi-domain proteins also tend to contain IURs. The reasons for this are less obvious and still debated (\citealt{DunkerEtAl2005}; \citealt{FongAndPanchenko2010}). IURs can serve as inter-domain linkers that hold multiple domains together in a protein (\citealt{Tompa2002}; \citealt{Tompa2005}). IURs can also occur in or near the binding sites of proteins, and may facilitate protein interactions (\citealt{DunkerEtAl2005};\citealt{UverskyAndDunker2010}). This would make IURs particularly relevant to multi-domain proteins because such proteins tend to have multiple interaction partners (\citealt{TordaiEtAl2005}; \citealt{EkmanEtAl2007}). Indeed, the need to interact with multiple partners is one reason why some proteins contain multiple domains (\citealt{Patthy2003}; \citealt{TordaiEtAl2005}; \citealt{TanEtAl2005}). Each domain can facilitate interactions with different partners.

	As stated previously, \cite{OlzschaEtAl2011} found that aggregation-prone proteins tend to be large and tend to contain IURs, which are characteristics of multi-domain proteins. Large proteins are more aggregation-prone than smaller proteins because they take longer to fold (or refold) into their native conformations, which increases the exposure of aggregation-prone amino acid sequences that are typically buried within the interior of the folded protein (\citealt{NetzerAndHartl1997}; \citealt{GoldschmidtEtAl2010}; \citealt{OlzschaEtAl2011}). This would be true for both recently synthesized proteins and for proteins that denature due to an environmental stress.  Large proteins also potentially contain more binding sites than smaller proteins, which would increase the likelihood for an oligomerization reaction to occur when large proteins approach each other (e.g. AndersenEtAl2010). Multi-domain proteins can also aggregate via ``domain swapping.'' This occurs when two domains in a multi-domain protein molecule are supposed to bind to each other, but instead they bind to the domains on another molecule (\citealt{NelsonAndEisenberg2006}; \citealt{RousseauEtAl2012}).

	Proteins that contain IURs tend to be aggregation-prone because the conformational flexibility of IURs allows them to align properly to form the cross-$\beta$ sheet structure that holds aggregates together (\citealt{CarrioEtAl2005}; \citealt{NelsonEtAl2005}; \citealt{Ventura2005}; \citealt{WangEtAl2008}; \citealt{OlzschaEtAl2011}; \citealt{RamshiniEtAl2011}; \citealt{StroudEtAl2012}). \cite{ZhangEtAl2013} have proposed that making a distinction between semi-disordered and completely disordered proteins is useful, as only the former are aggregation prone. Hence, it is important to realize that aggregation-prone proteins contain intrinsically unstructured regions (IURs), not that the entire protein is intrinsically disordered. Nevertheless, several amyloid-associated diseases are due to polypeptides that contain IURs (\citealt{UverskyEtAl2008}; \citealt{UverskyEtAl2009}; \citealt{BabuEtAl2011}). In addition, proteins containing IURs are metastable and are often in an unfolded state. As a consequence, proteins containing IURs are thought to have relatively short life-spans and rapid turnovers because they are continuously degraded by each cell's proteolytic machinery (\citealt{WrightAndDyson1999}).

	The abundance of large and IUR-containing proteins varies significantly across the domains of life. On the whole, prokaryotes produce shorter proteins than eukaryotes. About 65\% of prokaryote proteins are multi-domain whereas 80\% of eukaryote proteins are multi-domain (\citealt{ApicEtAl2001}). Furthermore, the median length of eukaryote proteins is 50\% longer than the median length of prokaryote proteins (\citealt{BrocchieriAndKarlin2005}). The relative abundance of proteins containing IURs are 2\%, 4\%, and 33\% for archaea, bacteria, and eukaryotes, respectively (\citealt{WardEtAl2004}). These differences have caused prokaryotes and eukaryotes to process their proteins differently. For example, prokaryotes typically translate proteins at a rate of 10-20 amino acids per second whereas eukaryotes typically translate proteins at a rate of 3-8 amino acids per second (\citealt{SillerEtAl2010}). As a consequence, most protein folding is post-translational in prokaryotes but co-translational in eukaryotes (\citealt{NetzerAndHartl1997}). Eukaryotes also have more complex chaperone systems that assist with the proper folding of their \textit{de novo} and stress-denatured proteins (\citealt{AlbaneseEtAl2006}). Thus, protein folding is more elaborate in eukaryotes than prokaryotes because their proteins are more aggregation prone. The truncation selection curves depicted in Figure \ref{fig:TruncationSelection} should be shifted to higher heterozygosity values for organisms that synthesize more aggregation-prone proteins, and indeed, heterozygosity is essentially a eukaryotic phenomena. There are no heterozygous bacteria and archaea. This argument will be expanded later to explain the haploid-diploid transition.

	A species' geographic distribution is affected by both environmental stress and its proteins' susceptibility to aggregation. Both \cite{KooninEtAl2002} and \cite{BrocchieriAndKarlin2005} proposed that hyperthermophiles are common in the Domain Archaea because the species in this domain produce short proteins that are not prone to aggregation. However, the idea can be extended to all prokaryotes that live in extreme environments. Thermophiles can be found among both the Bacteria and the Archaea. Additionally, both domains of life contain species that are able to grow at extremely high salinities (\citealt{KunteEtAl2002}), withstand desiccation (\citealt{Potts1994}; \citealt{Alpert2006}), or grow in freezing conditions (\citealt{Russell1998}). Not all prokaryotes can grow in extreme environments because additional adaptations, such as those promoting membrane integrity and DNA stability, are required (\citealt{KoningsEtAl2002}; \citealt{TrivediEtAl2005}). However, prokaryotes should be ``pre-adapted" to these extreme environments because they are not removed by truncation selection when they migrate into them. In contrast, truncation selection could prevent organisms that synthesize more aggregation-prone proteins, including many eukaryotes, from migrating into these environment (because their relative fitness would be zero), thereby limiting their distribution to less harsh environments.

\section{Ploidy Level} \label{sec:Ploidy}

\subsection{Environmental Stress and Polyploidy} \label{subsec:Polyploid}

	If the expression of two allozymes can inhibit the formation of soluble oligomers, then the expression of three or more allozymes should further inhibit their formation. Equation \ref{eq:GeneralOligomerRate} predicts that polyploid organisms should have lower rates of protein aggregation than diploid organisms. In addition, many polyploids are hybrid species (allopolyploids), and are heterozygous at more gene loci than diploids (\citealt{OttoAndWhitton2000}). However, even autopolyploids can express more than two allozymes at a given gene locus. Therefore, polyploid organisms should be able to tolerate more severe physical stresses than diploids, and in fact, may be ``pre-adapted'' to harsh environments in which diploid organisms cannot survive due to truncation selection. Indeed, since polyploidy and hybridization may be accompanied by fitness costs (such as high genetic loads, slower growth rates, and outbreeding depression), polyploid species may be restricted to harsh environments where they do not have to compete with diploid species (\citealt{OttoAndWhitton2000}).

	The occurrence of polyploid plants at high latitudes and altitudes was first observed in the 1940's (\citealt{Stebbins1950}; \citealt{Stebbins1984}), and recent research has confirmed that polyploid plants and animals frequently occur in frozen environments (\citealt{BeatonAndHebert1988}; \citealt{AdamowiczEtAl2002}; \citealt{BrochmannEtAl2004}; \citealt{LundmarkAndSaura2006}; \citealt{AguileraEtAl2007}; \citealt{OttoEtAl2007}; \citealt{AdolfssonEtAl2009}). \cite{BrochmannEtAl2004} analyzed data from the \textit{Pan-Arctic Flora (PAF) Checklist} (\citealt{ElvenEtAl2003}) and found that 73.7\% of arctic plants are polyploid. Plants in the most northerly arctic zone were hexaploid on average. In addition, 39.2\% of species within this zone were 7-ploid or higher, and 17.8\% of the species were 9-ploid or higher. However, since these plants reproduce primarily through self-fertilization, the heterozygosity of these plants is about half what their ploidy level would indicate.

	Polyploid plants are also positively associated with arid zones and deserts (\citealt{Spellenberg1981}; \citealt{RossiEtAl1999}; \citealt{HunterEtAl2001}; \citealt{PannellEtAl2004}; \citealt{JolyEtAl2006}; \citealt{SchuettpelzEtAl2008}). \cite{SenockEtAl1991} and \cite{HaoEtAl2013} showed that the ploidy level of \textit{{A}triplex canescens} increases in regions of the Chihuahuan Desert with increasing drought stress. Most resurrection plants, which grow in deserts and are capable of withstanding very high levels of desiccation, are polyploid (\citealt{BartelsAndSalamini2001}; \citealt{RodriguezEtAl2010}). Several studies have shown that a plant's drought tolerance increases with its ploidy level (\citealt{AlHakimiEtAl1998}; \citealt{XiongEtAl2006}). For example, \cite{Ramsey2011} compared the drought tolerances of hexaploid and tetraploid individuals belonging to \textit{Achillea borealis} and found the hexaploids were more tolerant. Furthermore, Ramsey compared the drought tolerances of newly formed hexaploid \textit{A. borealis} individuals to hexaploid individuals collected from the wild, and determined that a third of the the drought tolerance was achieved via genome duplication alone rather than adaptation. Another interesting example of polyploid adaptation to water stress may be the redwood, \textit{Sequoia sempervirens}, which is hexaploid and must cope with water stress due to its extreme height (\citealt{AhujaEtAl2002}; \citealt{KochEtAl2004}; \citealt{OldhamEtAl2010}). Polyploid animals are also positively associated with arid zones. For example, polyploid lizards are found in deserts (\citealt{TocidlowskiEtAl2001}; \citealt{Kearney2003}) and so is the only known polyploid mammal (\citealt{GallardoEtAl1999}; \citealt{GallardoEtAl2004}; \citealt{SvartmanEtAl2005}; \citealt{GallardoEtAl2006}; \citealt{TetaEtAl2014}).

	Polyploidy should also be associated with salinity stress because many species change their expression of \textit{Hsps} in both hypersaline and hyposaline environments (\citealt{Chang2005}; \citealt{DownsEtAl2009}; \citealt{TineEtAl2010}; \citealt{MonariEtAl2011}). However, evidence for a link between polyploidy and salinity is unfortunately tenuous. Polyploid plants appear to have greater tolerance to salt stress than their diploid relatives (\citealt{TalAndGardi1976}; \citealt{ShannonAndGrieve1999}; \citealt{AshrafEtAl2001}; \citealt{KumarEtAl2009}). Also, several species of polyploid brine shrimp, \textit{Artemia,} have been identified (\citealt{BrowneAndBowen1991}; \citealt{AmatEtAl2007}). Several papers have suggested that the radiation of polyploid \textit{Artemia} is related to the Messinian salinity crisis (e.g. \citealt{AghEtAl2007}), but other papers have disputed this claim (\citealt{BaxevanisEtAl2006}). Various papers have found that parthenogenic \textit{Artemia} tolerate both higher and lower salinities than sexuals, but none of these papers distinguish between polyploid and diploid parthenogens (\citealt{BrowneAndMacDonald1982}; \citealt{ZhangAndKing1993}; \citealt{ElBermawiEtAl2004}; \citealt{AghEtAl2007}). The distribution of polyploid \textit{Artemia} may instead be driven by latitude (\citealt{ZhangAndLefcort1991}).

	The geographical distribution of polyploid organisms suggests that they have an advantage in environments that promote protein aggregation. These trends can be understood in terms of a truncation selection model. Truncation selection would prevent individuals with low heterozygosities from migrating into harsh environments, thereby preventing some diploid species from expanding into harsh environments and adapting to them. In contrast, polyploid individuals may be partially pre-adapted to harsh environments because their heterozygosities are already sufficiently high to avoid truncation selection. Upon time, the polyploids will further adapt as favorable alleles increase in frequency. A dramatic illustration of this may be given in \cite{FawcettEtAl2009}, which argues that polyploids may have preferentially survived the Cretaceous-Paleogene mass extinction event.

\subsection{Organism Complexity, Protein Interaction Networks, and the Haploid/Diploid Transition} \label{subsec:HaploidDiploid}

	Protein aggregation may also potentially explain why some species are haploid and others are diploid. The most suggestive evidence for this is the relative stress tolerances of haploid and diploid organisms. Animals, for instance, have lower thermotolerances than bacteria and archaea, and some fungi species (\citealt{Portner2001}; \citealt{SalarAndAneja2007}). Even the Pompeii worm, which grows in hydrothermal vents, can not withstand temperatures greater than 55$^{\circ}$C for 2 hours (\citealt{RavauxEtAl2013}). Similarly, animals are less tolerant to desiccation stresses than prokaryotes and fungi, with the notable exceptions of bdelloid rotifers and tardigrades\footnote{Polyploidy may explain these exceptions. Bdelloid rotifers are descended from a tetraploid ancestor, and most freshwater and terrestrial tardigrades are polyploid (\citealt{Bertolani2001}; \citealt{HurEtAl2009})} (\citealt{Alpert2006}). The general trend appears to be that animals are less stress tolerant than fungi, which in turn are less stress tolerant than bacteria and archaea. Considering that the ability to produce \textit{Hsps} appears to control the upper thermal tolerances of organisms (\citealt{Portner2001}; \citealt{DillyEtAl2012}), the relative stress tolerances of animals, fungi, bacteria, and archaea may reflect the relative levels of protein aggregation that they must cope with.

	The trend in stress tolerance seems to match a trend in the ploidy levels of these organisms. Bacterial and archaeal species are haploid, animal species are diploid, and fungal species may either be haploid or diploid. Thus, bacterial and archaeal species may never have to cope with truncation selection for heterozygosity while all animal species may experience such truncation selection, even in relatively mild conditions. Fungal species may or may not experience truncation selection for heterozygosity depending on the environmental conditions or the stage in their life-cycle (see below). An underlying mechanism may generate the different levels of protein aggregation in these organisms, which may explain their relative stress tolerances and their ploidy levels.

	But why do different species experience different levels of protein aggregation? One explanation is that some species produce more aggregation-prone proteins than others. \cite{OlzschaEtAl2011} found that aggregation-prone proteins in human cells tend to be large and tend to contain IURs, which are characteristics of multi-domain proteins (\citealt{DunkerEtAl2005}; \citealt{FongAndPanchenko2010}). These proteins often have numerous interaction partners and are often involved in signal transduction and regulatory processes (\citealt{DunkerEtAl2005}; \citealt{WarringerAndBlomberg2006}; \citealt{UverskyAndDunker2010}; \citealt{OlzschaEtAl2011}). Such proteins facilitate organism complexity because they have important roles in coordinating and regulating the biochemical activities within and between cells, which is necessary for the development of complex organisms (\citealt{RubinEtAl2000}; \citealt{Patthy2003}; \citealt{DunkerEtAl2005}; \citealt{TanEtAl2005}; \citealt{TordaiEtAl2005}; \citealt{EkmanEtAl2007}; \citealt{UverskyAndDunker2010}). For example, \cite{StumpfEtAl2008} found that the human protein interaction network (PIN) contains approximately 650,000 protein interactions while the \textit{S. cerevisiae} PIN contains approximately 25,000-35,000 protein interactions.

	PINs are important for communication and coordinating all of the activities that occur in cells (\citealt{BarabasiAndOltvai2004}; \citealt{VinayagamEtAl2014}). Communication usually occurs through signaling cascades involving large numbers of proteins (\citealt{PawsonAndNash2000}; \citealt{BreitkreutzEtAl2010}; \citealt{VinayagamEtAl2011}). Coordination can involve the switching on or off of proteins via phosphorylation, transcription factors, proteolysis, etc. (\citealt{LeMosyEtAl2001}; \citealt{Walhout2006}; \citealt{Lopez-OtinAndHunter2010}; \citealt{WangAndChen2010}; \citealt{ChengEtAl2011}). Thus, the cell needs a sort of managerial class of proteins whose specialty is interacting with other proteins. These proteins tend to be larger than the median protein because they need multiple domains to facilitate all of their interactions (\citealt{TanEtAl2005}; \citealt{TordaiEtAl2005}; \citealt{EkmanEtAl2007}). The managerial proteins also tend to be metastable (or semi-disordered) which allows them to fold into multiple conformations that can bind to different partners. This flexibility is facilitated by IURs (\citealt{DunkerEtAl2008}). The managerial proteins appear to be prone to aggregation, and the molecular chaperone {Hsp90} appears to specializing in preventing such aggregation of this class of proteins (\citealt{Picard2002}; \citealt{SangsterEtAl2004}; \citealt{VabulasEtAl2010}).

	The proteins that facilitate complex PINs tend to be large and multi-domain because multiple domains are necessary to facilitate different interactions (\citealt{RubinEtAl2000}; \citealt{TanEtAl2005}; \citealt{TordaiEtAl2005}; \citealt{EkmanEtAl2007}; \citealt{ZmasekAndGodzik2011}). For example, \cite{XiaEtAl2008} found that the average number of interaction domains per protein increases with the number of cell types in an organism. Likewise, \cite{WangEtAl2005} found that proteins shared by \textit{S. cerevisiae}, \textit{D. melanogaster}, and \textit{H. sapiens} are similar in length; but proteins found in \textit{D. melanogaster} and \textit{H. sapiens,} but not in \textit{S. cerevisiae}, are on average 22\% longer than proteins shared by all three species. Finally, \cite{WarringerAndBlomberg2006} found that \textit{S. cerevisiae} proteins longer than 770 amino acids have more interaction partners, on average, than shorter proteins. Particularly abundant among these large proteins were transport proteins, proteases, kinases, and other signaling proteins. These proteins are responsible for signal transduction and regulating biochemical pathways (\citealt{SoporyAndMunshi1998}; \citealt{ManningEtAl2002}; \citealt{Lopez-OtinAndHunter2010}). In multicellular organisms, these proteins play vital roles in intercellular communication, regulation of the cell cycle, and cellular differentiation (\citealt{LeMosyEtAl2001}; \citealt{Schaller2004}; \citealt{Turk2006}; \citealt{VanderHoorn2008}; \citealt{KeshetAndSeger2010}). Thus, large proteins play an import role in coordinating and regulating the biochemical activities in organisms, and such proteins appear to be more abundant in complex organisms (\citealt{TordaiEtAl2005}; \citealt{EkmanEtAl2007}).

	Proteins containing IURs also facilitate complex PINs. IURs can serve as flexible inter-domain linkers that allow domains to move freely with respect to each other (\citealt{Tompa2002}; \citealt{Tompa2005}). These inter-domain linker regions are not merely structural, but often times serve as binding sites between proteins and their interaction partners (\citealt{BalazsEtAl2009}). IURs allow proteins to assume multiple conformations which in turn allows the proteins to bind to multiple partners (or allows multiple partners to bind to the same protein) (\citealt{KriwackiEtAl1996}; \citealt{TompaEtAl2005}; \citealt{OldfieldEtAl2008}; \citealt{TyagiEtAl2009}; \citealt{Bustos2012}). IURs may also speed up binding reactions via a ``fly-casting'' mechanism (\citealt{ShoemakerEtAl2000}). Finally, IURs provide easily accessible sites for post-translational modifications to proteins, which makes them important for biochemical regulation (\citealt{DunkerEtAl2002}; \citealt{KurotaniEtAl2014}). As a consequence of these structural properties, up to 94\% of transcription factors and more than 70\% of signaling proteins in eukaryotes contain IURs (\citealt{IakouchevaEtAl2002}; \citealt{LiuEtAl2006}; \citealt{UverskyAndDunker2010}). In short, IURs are important mediators of protein interactions, and proteins containing IURs are responsible for most signal transduction and biochemical regulation in eukaryotes.

	Thus, it appears that the proteins that facilitate organism complexity are also the proteins that make complex organisms more sensitive to environmental stresses. This can explain why higher ploidy levels are associated with complex organisms. Bacteria and archaea never experience truncation selection for heterozygosity, even in harsh environments. In contrast, animals must be diploid, even in mild environments, because they produce an abundance of proteins with numerous interaction partners. These proteins are typically large and typically contain IURs, so they tend to be aggregation-prone. As a consequence, animals must be heterozygous in order to be on the right side of the truncation curve (Figure \ref{fig:TruncationSelection}), and that requires them to be at least diploid. Organisms of intermediate complexity, such as plants and fungi, may be either haploid or diploid, depending on the species. Whether such species are haploid or diploid will depend on the abundance of aggregation-prone proteins that they produce and on the stresses to which their proteins are exposed. If this hypothesis is true, then the diploidy of complex organisms is largely necessitated by the physical constraints imposed by PIN complexity. It may be the case that the thermodynamic benefits of heterozygosity (Subsection \ref{subsec:Thermodynamics}) increase with an organism's complexity.

\subsection{Developmental Homeostasis} \label{subsec:Homeostasis}

	The hypothesis presented in the previous subsection shares several connections with the pioneering work from early metabolic heterosis theorists, such as I. Michael Lerner. \cite{Lerner1954} found that inbred plants and animals exhibited more morphological variability than outbreds, which Lerner attributed to a decline in developmental homeostasis (or developmental stability). In other words, less heterozygous individuals display greater degrees of aberrant growth and development, which result in morphological imperfections such as bilateral asymmetry. Over the years, other researchers have corroborated Lerner's findings (\citealt{RobertsonAndReeve1952}; \citealt{Eanes1978}; \citealt{Mitton1978}; \citealt{Soule1979}; \citealt{Mitton1995}), which in turn lead to the studies that found heterozygosity correlates with growth rate, metabolic efficiency, and protein turnover (\citealt{SinghAndZouros1978}; \citealt{ZourosEtAl1980}; \citealt{KoehnAndShumway1982}; \citealt{GartonEtAl1984}; \citealt{MittonAndGrant1984}; \citealt{HawkinsEtAl1986}; \citealt{HawkinsEtAl1989}). Thus, the field of metabolic heterosis studies can trace its roots back to Lerner's work on developmental homeostasis.

	The hypothesis presented in this subsection can explain how low heterozygosity would disrupt developmental homeostasis. The proteins that regulate proper growth and development are also the proteins that are most susceptible to aggregation. Both environmental stress and inbreeding can lead to elevated rates of aggregation for signaling and regulatory proteins, which could potentially disrupt intercellular communication, regulation of the cell cycle, proper cell differentiation, etc. The cumulative result would be a disruption in developmental homeostasis as described by I. Michael Lerner. Thus, protein aggregation can potentially explain much of the phenomena that concerned early researchers working with allozymes (see \citealt{Mitton1997} for review.)

\subsection{Plants and Alternation of Generations} \label{subsec:Plants}

	The trends described in the previous subsections hold for plant species as well. Plants typically alternate between a haploid gametophyte generation and a diploid sporophyte generation. However, the dominant generation varies between divisions. For example, bryophytes have a dominant gametophyte generation and a short-lived sporophyte generation while spermatophytes are primarily diploid (the sporophyte generation is dominant). In addition, ferns have independent gametophyte and sporophyte generations. The theory presented in Subsection \ref{subsec:HaploidDiploid} predicts that fewer aggregation-prone proteins should be produced by the haploid gametophytes than in the diploid sporophytes. Two types of circumstantial evidence support this prediction.

	First, there is a relationship between complexity and ploidy level. The haploid-dominant bryophytes are relatively simple plants, typically 2 cm tall and one cell thick. The diploid-dominant spermatophytes are complex and include flowering plants. The ferns alternate between a haploid generation that is simple, resembling bryophytes, and a diploid generation that is significantly larger and more complex. Thus, according to the hypothesis presented in the previous subsection, spermatophytes and fern sporophytes should have more complex PINs and should synthesize more aggregation-prone proteins than bryophytes and fern gametophytes.

	The second type of evidence is the relative stress tolerances of the different plant divisions. The relationship between ploidy level and stress tolerance in plants is similar to the trend described in Subsection \ref{subsec:HaploidDiploid} for animals, fungi, and prokaryotes. Bryophytes are much more tolerant of freezing, desiccation, and salinity stresses than spermatophytes (\citealt{Alpert2000}; \citealt{OliverEtAl2005}; \citealt{WangEtAl2009}; \citealt{GaffAndOliver2013}). Their ability to tolerate such stresses is comparable to lichens, and they can be found, along with lichens, in extremely cold and arid environments not inhabited by more complex plants (\citealt{Longton1988}; \citealt{Alpert2006}; \citealt{ProctorAndTuba2002}; \citealt{KrannerEtAl2008}). The desiccation tolerance of ferns is more complex. Fern sporophytes are comparable to spermatophytes in their ability to tolerate desiccation, but fern gametophytes are comparable to bryophytes (\citealt{WatkinsEtAl2007}; \citealt{Hietz2010}). In fact, some tropical fern species have lost the sporophyte stage of their life cycle and now exist as asexually reproducing gametophytes, which has allowed them to migrate into colder and drier habitats than their sporophyte-producing relatives (\citealt{Farrar1978}; \citealt{Farrar1990}). Thus, bryophytes and fern gametophytes probably express fewer aggregation-prone proteins than spermatophytes and fern sporophytes. This would explain the relative order of plant stress tolerances: bryophyte $\approx$ fern gametophyte {\textgreater} spermatophyte $\approx$ fern sporophyte. The trend corresponds to the relative complexity of the plant divisions and to their ploidy levels.

	The above observations on relative stress tolerances should not be taken to imply that bryophytes are never subjected to truncation selection for heterozygosity. For example, allodiploid species of bryophytes (hybrids with two sets of chromosomes) have been identified, and they increase in frequency with latitude (\citealt{WyattEtAl1988}; \citealt{RiccaEtAl2008}). Bryophyte species probably produce few enough aggregation-prone proteins that they do not experience truncation selection for heterozygosity in mild environments, but they may produce enough aggregation-prone proteins that they experience truncation selection in harsher environments. Thus, bryophytes, ferns, and spermatophytes all exhibit ploidy level increases in harsh environments. The difference between bryophytes and spermatophytes is that bryophytes have a haploid chromosome set baseline whereas spermatophytes have a diploid chromosome set baseline. Bryophytes may also experience truncation selection for heterozygosity in their sporophyte generations. Like ferns, the sporophyte generations of bryophytes are diploid and cannot survive in as harsh of environments as the gametophyte generations (\citealt{StarkEtAl2007}). This may indicate that the sporophyte generation of bryophytes express aggregation-prone proteins that are not produced by the gametophyte generation.

\subsection{Alternation of Generations and the\\ Evolution of Complexity} \label{subsec:Complexity}

	When comparing the life-cycles of green algae, bryophytes, ferns, and spermatophytes, there appears to be a progression from haploid dominant species, to species that alternate between haploid and diploid generations, to diploid dominant species. In other words, diploid dominant species did not evolve directly from haploid dominant species, but instead, from species that alternated between haploid and diploid generations. The truncation selection model developed in this paper can explain this evolutionary sequence.

	Recall that a species will go extinct if all individuals are on the wrong side of a truncation curve, and that diploid species produce aggregation-prone proteins that contain multiple domains and IURs. These two statements imply that it would be impossible for strictly haploid species to produce an abundance of aggregation-prone proteins because truncation selection would guarantee their extinction. Hence, the need for diploidy would never arise. Diploid individuals are unlikely to successfully compete against haploid individuals if there is no immediate advantage to diploidy, especially if there are advantages to haploidy as described in the next subsection.

	However, a spore-producing generation can circumvent this barrier by providing an immediate advantage to diploidy. Sexual reproduction requires haploid dominant species to possess at least a temporary diploid generation. For such organisms, sexual reproduction is often accompanied by the production of spores or cysts. This may facilitate inbreeding avoidance since spores often serve as a means of dispersal, or a mixed cytoplasm may increase the ability of a spore or cyst to tolerate environmental stresses (Equation \ref{eq:TruncationSelection}). A mixed cytoplasm may also increase the longevity of a spore or cyst since it would slow down the accumulation of soluble oligomers in the dormant organism over time. For fungi, there is some evidence that sexual spores are more stress tolerant than asexual spores\footnote{The spores in these studies were haploid, but they could have inherited a mixed cytoplasm from their diploid parent cells during meiosis.} (\citealt{GrishkanEtAl2003}; \citealt{Dijksterhuis2007}; \citealt{Trapero-CasasAndKaiser2007}). Regardless, sexual reproduction requires organisms to possess at least a diploid zygote in their life-cycle, and this often accompanies spore production. Some species have evolved a separate diploid (or dikaryotic) spore-producing generation, such as sporophytes in plants or ascocarps and basidiocarps in dikaryotic fungi, because it aids with spore dispersal. This spore-producing generation has the potential to evolve complexity over time because truncation selection would not prevent it from expressing genes that encode the aggregation-prone proteins that facilitate complexity. In contrast, truncation selection would prevent strictly haploid spore-producing species from becoming more complex over time.

	Given the constraints that truncation selection would impose on haploid organisms, the evolution of diploid-dominant organisms may have proceeded along the following steps: 1) Bacterial and archaeal species produce proteins that are not susceptible to aggregation, so they do not have a heterozygous diploid stage in their life-cycle, even when producing spores; 2) some eukaryotic species (e.g. some algal and fungal species) produce enough aggregation-prone proteins that their spores benefit from a mixed cytoplasm, which requires at least a temporary diploid stage in their life-cycle\footnote{This could have evolved when cells from closely related species fused together to form a diploid cell, which would be analogous to the formation of polyploid species via hybridization}; 3) Some species have longer-lived diploid, spore-producing generations in their life-cycles, perhaps because they facilitate spore dispersal (e.g. bryophytes) 4) the diploid stages of some species' life-cycles have become more complex over time because truncation selection has not prevented them from producing aggregation-prone proteins with numerous interaction partners (e.g. ferns, dikaryotic fungi); and 5) the haploid life-cycle stage has become temporary in some multicellular species, which has resulted in diploid-dominant species (e.g. animals and spermatophytes). Thus, the evolution of diploid-dominant species would be a very gradual process, unlike the evolution of polyploidy, which occurs in a single generation.

	The freshwater green algae Charales may provide support for the above evolutionary sequence. Charales are exclusively haploid, lacking a sporophyte stage in their life-cycle (\citealt{BeckerAndMarin2009}). They are also relatively complex compared to other green algae, but are still much simpler than angiosperms (\citealt{Lee2008}). However, \cite{GrahamAndGray2001} argues that Charales ``are not competitive with freshwater angiosperms and rarely share freshwater habitats with them.'' In fact, the fossil record shows that Charales's species diversity has declined over time since the appearance of angiosperms, and that Charales may be an evolutionary dead end (\citealt{GrahamAndGray2001}). Given that Charales species are closely related to embryophytes, and that complexity seems to have given angiosperms a competitive advantage over Charales species, it is natural to ask why Charales species did not become more complex over time. Competition between Charales individuals should have lead to an increase in complexity over time, just as competition favored the evolution of complex angiosperms.

	The absence of a sporophyte stage in the Charales life-cycle may explain their relatively simple morphology since, as argued in the previous four paragraphs, truncation selection would impose an upper limit on the abundance of aggregation-prone proteins that the haploid Charales species can produce. This in turn would impose an upper limit on the complexity of Charales species because proteins with numerous interaction partners, which are required in complex organisms, tend to be aggregation-prone. Thus, Charales species may have hit an evolutionary dead end because they do not possess a sporophyte generation in their life-cycle. In contrast, an upward-growing sporophyte generation is beneficial to embryophytes because it facilitates spore dispersal through the air. The possession of a diploid sporophyte generation may have removed a barrier to the evolution of complexity in some embryophyte lineages, which has allowed the spermatophytes to become large, complex organisms.

\subsection{The Advantages of Haploidy} \label{subsec:Haploid}

	The theory developed in Subsection \ref{subsec:HaploidDiploid} attempts to explain why diploidy is advantageous. However, the occurrence of organisms that alternate between haploid and diploid life-cycles stages (e.g. ferns and \textit{Ulva}) suggests that haploidy has advantages. Such organisms could stay permanently diploid (with brief haploid stages for the purpose of sexual reproduction) if there was no advantage to haploidy. Two such potential advantages are genetic load and growth rate.

	Genetic load should favor the evolution of lower ploidy levels (\citealt{MableAndOtto1998}). The genetic load in a population is directly proportional to the mutation rate (\citealt{Haldane1937}). Thus, if mutation rates are relatively constant at all ploidy levels, then a population of diploid individuals should have twice the genetic load of a population of haploid individuals (\citealt{OttoAndWhitton2000}, \citealt{GersteinAndOtto2009}). As a consequence, populations of haploid individuals should have higher mean fitnesses than populations of otherwise identical diploid individuals (\citealt{MableAndOtto1998}).

	Higher ploidy levels are also disadvantageous because they lead to slower growth rates. This has been observed in polyploid plants, which typically grow and mature more slowly than their diploid relatives (\citealt{OttoAndWhitton2000}; \citealt{HessenEtAl2009}). Also, diploid gametophyte lines of one bryophyte species grow $\approx$70\% as fast as haploid lines on full medium (\citealt{SchweenEtAl2005}). Thus, haploidy might be beneficial to organisms that face strong selection for high growth rates. Haploidy would also be particularly beneficial to single-celled organisms, in which cell division rates are directly tied to fecundity.

	The growth rate hypothesis is particularly useful when trying to understand the life cycles of plants. All plants must compete for limited space (\citealt{GurevitchEtAl1990}; \citealt{GremerEtAl2013}), so a faster growth rate would give haploid plants an advantage over diploid plants when attempting to compete for access to land. For instance, bryophytes reproduce asexually via fragmentation and sexually via spore production. In both cases, the gametophyte plants must quickly grow from only a few cells in order to establish themselves in a partition of land. Ferns also disperse themselves via spores made up of only a few cells. Their simple haploid gametophyte generations may also be beneficial for competition over access to land. Once established, the ferns reproduce sexually and produce their complex, diploid sporophyte generations, which don't have to compete for access to land because they grow out of their gametophyte parents. In contrast, spermatophytes disperse themselves via seeds that carry entire diploid plant embryos. The plant embryos can quickly establish themselves in a partition of land, despite a slower growth rate, because they are already partially developed prior to germination. This might allow spermatophytes to be diploid, complex organisms for the bulk of their life-cycle, which in turn might allow them to utilize complex structures (e.g. flowers) for all of their life processes, including sexual reproduction.

\section{Conclusion} \label{sec:Conclusion}

	This paper attempts to provide a biochemical basis for heterosis and selection for heterozygosity. It then shows how individuals with higher heterozygosities are favored with increasing environmental harshness and organism complexity, which will also favor higher ploidy levels. In addition, the hypothesis was used to explain the different life-cycles of plants and the results of numerous experiments that have found heterozygosity correlates with reduced protein turnovers and higher metabolic efficiencies. Thus, the hypothesis can explain numerous field observations and experimental data in which heterozygosity and ploidy level are variables. Future research should be able to establish whether heterozygous advantage has a thermodynamic basis, and whether organism complexity and environmental harshness are in fact determinants of each species ploidy level.

	The hypotheses developed in this paper have important implications for breeding more stress tolerant varieties of crops. Several algorithms have been developed for identifying aggregation-prone amino acid sequences in proteins (\citealt{TartagliaEtAl2008}; \citealt{GoldschmidtEtAl2010}). These algorithms can be used to identify the proteins that are susceptible to aggregation, which in turn may identify gene loci where heterozygosity is most beneficial. Such identification may prove helpful in developing crop varieties that can withstand the physical stresses caused by global climate change (\citealt{LobellEtAl2013}). There is already some evidence that global climate change is leading to selection for heterozygosity in animal populations (\citealt{ForcadaAndHoffman2014}). Identifying aggregation-prone proteins may also be helpful in further improving crop yields as has occurred throughout the 20$^{th}$ century.

\section*{Acknowledgments}

I would like to thank Jeffry Mitton and Anthony J. Hawkins for their recommendations to improve this paper and encouragement. I would especially like to thank Dr. Mitton for his thorough review of the manuscript.

\nocitepaper{*}  
\bibliographypaper{ploidy3} 

\newpage
\clearpage 

\appendix
\section {} \label{sec:Appendix}

The purpose of this appendix is to highlight some applications of the theory presented in the main text in light of some papers published since the text's publication. The applications involve the evolution of syngamy, polygenic adaptation, and the evolution of sexual reproduction. I think our understanding of each of these phenomena is enhanced by the theory of truncation selection for heterozygosity presented in the main text. I do not think the ideas presented here warrant a new paper at the present moment, but I think the ideas may serve as food for thought, which may benefit other theorists. I will start with the evolution of syngamy, which I regard to be the strongest section of this appendix. I will then move on to polygenic adaptation and the evolution of sexual reproduction, which is much more speculative.

\subsection{The evolution of syngamy} \label{sec:Syngamy}

One of the key ideas presented in the main text is that heterozygosity throughout an organism's genome confers resistance to physical stresses that promote protein aggregation. Figure \ref{fig:TruncationSelection}b depicts two truncation curves for different levels of stress. Individuals with sufficiently high heterozygosity will survive a given level of stress. In contrast, individuals with too low of a heterozygosity will have a very low, if not negligible, fitness.  The critical heterozygosity level depends on the stressfulness of the environment. The more stressful the environment, the higher the requisite heterozygosity for surviving in the environment will be. Hence, individuals with low heterozygosity will only survive in mildly stressful environments.

The relationship between heterozygosity and stress tolerance has important implications for spores. The main text briefly touched on this subject in Section \ref{subsec:Complexity}, but I would like to elaborate more here. Many fungal species produce both asexual spores, usually referred to as conidia, and sexual spores, which have many different names. For the purposes of simplicity, I will focus on ascospores and basidiospores for the sexual spores. There is ample evidence that sexual fungal spores are more stress tolerant than asexual spores (\citealt{Dijksterhuis2007}; \citealt{DijksterhuisEtAl2007}; \citealt{Trapero-CasasAndKaiser2007}). In fact, \cite{GrishkanEtAl2003} showed that the abundance of sexual spores increases relative to asexual spores along an aridity gradient in Israel, suggesting that that likelihood of fungal species producing sexual spores increases with aridity.

The theory presented in the main text provides a reason for the relative stress tolerance of fungal spores. Sexual spores are produced after meiosis has occurred and should possess a mixed cytoplasm. Typically, only the nucleus of the proto-spore cell undergoes meiosis. The four daughter nuclei then align, and barriers form between the nuclei to produce four separate spore cells (\citealt{Mauseth2009}; \citealt{Neiman2005}). Therefore, sexual spores can inherent a mixed cytoplasm that contains two different allozymes, even though the spore is technically haploid. In contrast, asexual spores are produced via mitosis of haploid parent cells and have no possibility of containing a mixed cytoplasm. Thus, sexual spores should have lower total rates of soluble oligomer formation than asexual spores, and thus greater stress tolerance (Sections 2 and 3 of the main text).

Spore longevity plays a role as well. \cite{HongEtAl1997} shows that the longevity of conidia decreases with temperature and desiccation stress. This would make sense if the toxicity of soluble oligomers played a key role in spore longevity. Suppose that spore viability followed an equation similar to Equation \ref{eq:TruncationSelection} from the main text. There would be some concentration of total soluble oligomers that would kill 50\% of the spores (represented as $C_{50}$ in Equation \ref{eq:TruncationSelection}). The concentration of soluble oligomers in the spore would be affected by both time and the rate at which the oligomers form. Spores in a stressful environment would reach the C50 concentration faster than spores in a milder environment. If possessing a mixed cytoplasm lowers the rate at which soluble oligomers accumulate, then sexual spores should have higher longevities than asexual spores under equivalent environmental conditions.

Everything that I have summarized in this section can now be used to develop a theory for the evolution of syngamy. Since sexual spores possess higher stress tolerances and longevities than asexual spores, they should be able to weather harsher environmental conditions. \cite{DijksterhuisEtAl2007} and \cite{GrishkanEtAl2003} provide evidence for the greater hardiness of sexual spores relative to asexual spores. Thus, the ability to produce sexual spores would provide an immediate advantage to possessing a mixed cytoplasm that contains different allozymes for a given protein. For haploid organisms, the easiest way to obtain a mixed cytoplasm is by fusing two cells with different cytoplasm contents together, a process usually called syngamy.

Thus, haploid organisms may fuse their cells with the cells from other individuals in order to obtain a mixed cytoplasm that enables the production of spores that can withstand harsh conditions. This benefit would apply not only to fungi, but also to other organisms possessing an extended haploid stage in their life-cycle, such as green algae and bryophytes. This would provide an immediate advantage to the evolution of syngamy in haploid organisms.

Since the benefits of syngamy only arise if the fusing cells contain different cytoplasm contents, the organisms undergoing syngamy must guarantee that their cytoplasm contents are different. This may underlie the evolution of mating systems in organisms with extended haploid life-stages. Most such organisms follow a simple +/- mating system, wherein half the population consists of (+)-cells and the other half consists of (-)-cells. Only cells of opposite sign can fuse together, thus decreasing the likelihood that the fusing cells contain similar cytoplasm contents. The green algae \textit{Chlamydomonas reinhardtii} is an example of a species with such a life cycle. However, some fungi have evolved quite elaborate mating systems to reduce the risk of relatives fusing together (\citealt{Kothe1996}).

Sibling avoidance may also affect the timing of meiosis itself. Some spores, such as zygospores and the spores of \textit{C. reinhardtii}, undergo meiosis after spore germination. Other spores, such as ascospores, undergo meiosis prior to spore formation. The former process would seem to guarantee that the spores contain a mixed cytoplasm since the spores are diploid, but the process would also guarantee that siblings occupy the same area since meiosis occurs after germination. On the other hand, the latter process would allow sibling spores to disperse separately, thereby reducing the odds that siblings germinate in the same area. Thus, the latter process would be beneficial for reducing the risk of siblings undergoing syngamy. However, the latter process would only work if meiosis can be carried out in such a way as to generate haploid spores with mixed cytoplasm contents (\citealt{Mauseth2009}; \citealt{Neiman2005}).

Finally, I want to discuss why cells started fusing together in the first place. Desiccation and desiccating agents, such as polyethylene glycol (PEG), are known to promote fusion of membranes by weakening hydrophobic interactions (\citealt{RoyEtAl2016}; \citealt{ErkutEtAl2012}; \citealt{PedrazzoliEtAl2011}; \citealt{HoekstraEtAl2001}; \citealt{MacDonald1985}). Thus, it is quite possible that cell fusion first occurred by accident when cells were exposed to desiccating conditions. Ironically, the cells that fused together were more likely to survive the desiccating conditions because their cytoplasm contents were mixed. The necessary allozyme diversity could have been provided by nearly neutral mutations that had accumulated in the population or by hybridization between closely related species. Later, cells would evolve mechanisms to make the fusion process less haphazard and more systematic, which would give rise to proper syngamy.

The resulting fusion process would be similar to the “parasexual” cycle that has been observed in the yeast \textit{Candida albicans}, which undergoes cell fusion under stressful conditions and increases its ploidy from diploid to tetraploid. \textit{C. albicans} does not undergo traditional meiosis in order to reduce its ploidy level. Instead, it destroys chromosomal material until the diploid state is restored, hence why the process is called “parasexual” (\citealt{SherwoodAndBennett2009}; \citealt{ForcheEtAl2008}; \citealt{BennettAndJohnson2003}). Organisms that alternate between haploid and diploid life-stages may have used a similar parasexual process prior to the evolution of crossing over, chromosome segregation, and proper meiosis.

Thus, the benefits of heterozygosity discussed in the main text may provide an evolutionary advantage to syngamy that is independent of recombination. The reduction of soluble oligomer formation would have provided an immediate evolutionary benefit to fusing cells. Hence, the evolution of syngamy to form hardier spores may have been a prerequisite step to the evolution of meiosis and proper sexual reproduction.

\subsection{Polygenic adaptation and the benefits of sexual reproduction} \label{sec:PolygenicSex}

Several authors have argued that adaptation, especially in humans, may be highly polygenic, resulting from allele frequency shifts across many gene loci (\citealt{Gnecchi-RusconeEtAl2018}; \citealt{BergeyEtAl2018}; \citealt{BoyleEtAl2017}; \citealt{BergAndCoop2014}; \citealt{HancockEtAl2010}; \citealt{PritchardAndDiRienzo2010}; \citealt{PritchardEtAl2010}). However, there is some debate over the likelihood of such adaptation and proper interpretation of the data (\citealt{HollingerEtAl2019}; \citealt{BergEtAl2019}). For the sake of argument, this section will assume that such adaptation occurs.

\cite{HollingerEtAl2019} extensively modeled the conditions under which polygenic adaptation is likely to occur. They concluded that there were only two such conditions. Either the background mutation rate for the trait must be very large (requiring that the trait is controlled by hundreds or thousands of gene loci), or genetic variation must be maintained by balancing selection. The authors rule out the possibility of the former condition since it is unlikely that many traits are controlled by so many gene loci, and the authors are skeptical of the second condition since it would require balancing selection to be operating on many gene loci. Therefore, \cite{HollingerEtAl2019} argues that adaptation typically occurs from some combination of hard and soft sweeps.

However, one of the goals of my paper was to argue that balancing selection could be operating throughout the genomes of complex organisms. The truncation selection for heterozygosity model developed in the main text could easily maintain polymorphisms at frequencies necessary for polygenic adaptation \cite{Wills1978}. Moreover, the model should allow any number of alleles to be maintained at a given gene locus, as long as they do not co-aggregate. The model also does not require selection to occur at any particular gene locus, as long as individuals in the population have a sufficiently high heterozygosity. This generality of the truncation selection model means that any number of alleles could be segregating within a population at elevated frequencies, thereby providing a substantial source of genetic diversity within the population.

An important implication emerges when the ideas from my paper are combined with \cite{HollingerEtAl2019}. If polygenic adaptation requires balancing selection, and balancing selection is mostly restricted to complex organisms, then polygenic adaptation should also be restricted to complex organisms. In contrast, simple organisms (especially prokaryotes) should primarily adapt through a combination of hard and soft sweeps (e.g. \cite{Barroso-BatistaEtAl2014} and \cite{BarrickAndLenski2009}). Thus, selective sweeps at a few gene loci likely play a larger role in prokaryotic adaptation, while polygenic adaptation via allele frequency shifts is likely more common in complex plants and animals (\citealt{RadwanAndBabik2012}). Note that the modes of adaptation are not dichotomous, instead they represent opposite ends of an adaptation spectrum. Hence, simple organisms adapt in a more sweep-like fashion whereas the adaptation of complex organisms is more ``polygenicish.''

If this theory is correct, then the level of genetic diversity is not the variable that causes simple and complex organisms to adapt differently. Instead, differences in the composition of standing genetic variation are what cause populations of simple and complex organisms to adapt differently. In the absence of balancing selection, the relative frequencies of alleles that constitute a population's standing variation are determined by mutation-selection-drift balance (\citealt{HollingerEtAl2019}). The relative allele frequencies will be highly heterogeneous, with a few alleles occurring at relatively high frequencies and many alleles occurring at very low frequencies. When the environment shifts, some of the alleles in the pool of standing variation will become beneficial. The few beneficial alleles that achieve high frequencies due to genetic drift are more likely to become established in the population (i.e. achieve a frequency that is less influenced by genetic drift, which allows further frequency changes to be more deterministic). As a result, only a few beneficial alleles will become established in the population at any given time and adaptation will be sweep-like. In contrast, balancing selection can homogenize the distribution of allele frequencies in the pool of standing genetic variation by elevating the frequencies of many alleles above the level maintained by mutation-selection-drift balance. This allows beneficial alleles at more gene loci to become established simultaneously when the environment shifts. Since alleles at many gene loci have reached frequencies that allow for more deterministic frequency changes, adaptation is more polygenic with balancing selection (\citealt{HollingerEtAl2019}). Therefore, balancing selection allows populations of complex organisms to adapt in a more polygenic manner than populations of simple organisms, even though the populations of complex organisms may contain smaller pools of standing genetic variation (due to smaller population sizes, for example).

My primary interest in this subject is not the mechanism of adaptation, per se, but rather its implications for the evolution of sexual reproduction. \cite{HickeyAndGolding2018} argues that sexual reproduction could be maintained in natural populations, even if asexually reproducing individuals have a two-fold reproductive advantage, when adaptation is polygenic. The authors support their hypothesis with computer simulations that generate results similar to those published in previous papers (\citealt{Roze2014}; \citealt{OttoAndBarton2001}). Hence, the hypothesis presented in \cite{HickeyAndGolding2018} is not wholly original, but is an update to a theoretical approach that extends all the way back to August Weismann, R. A. Fisher, and Hermann Muller (\citealt{HickeyAndGolding2018}).

The simulations in \cite{HickeyAndGolding2018} consider a population of organisms with a single chromosome containing 100 bi-allelic gene loci. At each gene locus, an allele occurring at 5\% frequency within the population is undergoing positive selection. The authors point out that the probability of a single chromosome containing the favored allele at 25 gene loci is less than one in a billion. Hence, no individual organism within the population will possess the favored allele at all 100 gene loci.

The results of the simulation show that each favored allele increases from 5\% to approximately 35\% on average. Thus, adaptation was occurring via allele frequency shifts. Furthermore, the sexually reproducing individuals outcompete the asexually reproducing individuals even though the asexual individuals possess a two-fold reproductive advantage over the sexual individuals. Therefore, sexual reproduction may be favored in species undergoing repeated polygenic adaptation.

Here is where my paper comes into the picture. Remember that the main text says protein aggregation becomes worse as organisms increase in complexity. This is because complexity is facilitated by a class of ``managerial proteins'' that have numerous interaction partners. The managerial proteins serve as hubs in complex protein interaction networks, coordinating various activities and keeping the regulatory and signal transduction machinery running in sync. Among the ranks of these managerial proteins are kinases, proteases, phosphatases, transcription factors, etc. (see main text for elaboration). The managerial proteins are typically large and multi-domain, which facilitates numerous protein interactions. The proteins also typically contain intrinsically unstructured regions (IURs) that allow the proteins to assume multiple conformations (they are metastable), which provides another method of facilitating numerous interactions. Unfortunately, these physical properties tend to make the managerial proteins prone to aggregation. Thus, complex organisms face a greater protein aggregation burden than simple organisms and are more likely to experience truncation selection for heterozygosity. For this reason, complex organisms are typically diploid, whereas simple organisms are typically haploid (see main text for more elaboration).

Combining the hypotheses of \cite{HollingerEtAl2019}, \cite{HickeyAndGolding2018}, and my paper together yields a potential explanation for why sexual reproduction is widespread among complex organisms, but rarer among simple organisms. Sexual reproduction is evolutionarily advantageous when adaptation occurs via allele frequency shifts at many, possibly hundreds of loci (\citealt{HickeyAndGolding2018}). However, this mechanism of adaptation requires balancing selection to be operating on many gene loci throughout the genome (\citealt{HollingerEtAl2019}). Such balancing selection only operates on organisms of sufficient complexity to experience a heavy protein aggregation burden (my paper). \textbf{\textit{Thus, sexual reproduction is widespread among complex organisms because they experience the balancing selection necessary to make polygenic adaptation work. In contrast, simple organisms typically reproduced asexually because they do not experience balancing selection and are more likely to adapt via selective sweeps.}}

It is almost as if simple and complex organisms are living in different worlds. Simple organisms do not experience truncation selection for heterozygosity, are haploid, and adapt via a combination of hard and soft sweeps. Complex organisms do experience truncation selection for heterozygosity, are diploid, and adapt via frequency shifts at many gene loci (polygenic adaptation). It would seem that genetic diversity is much more important for complex organisms than for simple organisms, and this explains why complex organisms are more likely to reproduce sexually.

Is the hypothesis true? I do not know. The hypothesis is highly speculative, but I think it does warrant some consideration because it may assist other theorists thinking about these matters, which is why I have included it in this appendix.

\subsection{Recombination and simple organisms} \label{sec:Phases}

Many critics would object to the claim that simple organisms do not reproduce sexually (e.g. \citealt{Vos2009}). There is no doubt that homologous recombination is widespread among bacteria, for example. However, there are important differences. Recombination in bacteria is not coupled to reproduction, and bacteria can go many generations without recombination. In contrast, animals generally undergo sexual reproduction every generation (though cyclical parthenogens are a notable exception). Bacterial recombination only modifies a short and localized sequence of DNA in the bacterial chromosome, whereas crossing over in eukaryotes typically involves recombination at multiple sites along the chromosome, resulting in more shuffling of genetic material. As a consequence, bacterial recombination usually reduces genetic diversity while sexual reproduction usually increases genetic diversity (\citealt{IranzoEtAl2019}; \citealt{AmburEtAl2016}; \citealt{GorelickAndHeng2010}). Therefore, I am partial to alternative explanations for bacterial recombination than those used to explain sexual reproduction in organisms that undergo a complete meiotic cycle.

My preferred explanation for bacterial homologous recombination is that it assists with DNA repair (\citealt{BernsteinEtAl2011}; \citealt{BernsteinEtAl1989}). \cite{BernsteinEtAl2011} summarizes many empirical and theoretical justifications for the DNA repair hypothesis. The most convincing are that Rec proteins play a role in both DNA repair and homologous recombination, and that homologous recombination seems to be upregulated in bacteria exposed to conditions that promote DNA damage.

If DNA repair is the primary function of homologous recombination in bacteria, and combining beneficial alleles is the primary function of crossing over in complex eukaryotes, then there must have been a transition in function during the evolution of eukaryotes. My current thinking on the matter is that the transition occurred over four phases:
\begin{enumerate}
\item Homologous recombination: Homologous recombination evolved in the prokaryotes because it plays an important role in DNA repair (\citealt{BernsteinEtAl2011}; \citealt{BernsteinEtAl1989}). Prokaryotes do not alternate between haploid and diploid generations because protein aggregation is not a substantial enough problem for them to benefit from heterozygosity, even under stressful conditions (main text). DNA repair probably reduces genetic diversity in prokaryotic populations (\citealt{IranzoEtAl2019}; \citealt{AmburEtAl2016}; \citealt{GorelickAndHeng2010}), but this may not significantly affect the ability of prokaryotes to adapt because their adaptation is typically driven by hard and soft sweeps rather than via polygenic adaptation, which relies more on standing genetic variation (see previous section).

\item Syngamy: Protein aggregation still does not impose much of a burden for simple eukaryotes, so they can survive with a prolonged haploid generation. However, protein aggregation is more burdensome under stressful conditions, and these eukaryotes survive such conditions by producing spores with a mixture of allozymes. The eukaryotes undergo syngamy to produce the needed mixed cytoplasm and, thus, are temporarily diploid (this appendix). The result is a parasexual cycle wherein extra DNA is loss by degradation analogous to what is seen in \textit{C. albicans} (\citealt{SherwoodAndBennett2009}; \citealt{ForcheEtAl2008}; \citealt{BennettAndJohnson2003}). DNA repair is still the primary function of homologous recombination.

\item Synapsis: A key step in the evolution of meiosis from mitosis is the synapsis of homologous chromosomes (\citealt{WilkinsAndHolliday2009}). DNA repair proteins seem to have been recruited to assist with the formation of chiasmata (\citealt{BernsteinEtAl2011}). The formation of chiasmata may have initially improved the fidelity of recombination (by preventing ectopic recombination for example), but the chiasmata also help homologous chromosomes to properly align, so they can segregate correctly during meiosis and ensure that each daughter cell has a proper karyotype (\citealt{BernsteinEtAl2011}; \citealt{GorelickAndHeng2010}; \citealt{WilkinsAndHolliday2009}; \citealt{Heng2007}). A true meiotic cycle exists by the time this phase is completed, which allows some eukaryotes to undergo systematic transitions between haploid and diploid life stages. This alternation between haploid and diploid life stages is widespread among fungi, algae, and plants.

\item Polygenic Adaptation: Some eukaryotes (e.g. animals) are primarily diploid. Meiosis would seem to be unnecessary for these species, which could reproduce solely through mitosis. Yet, meiosis has not been loss in these species because sexual reproduction is beneficial when species adapt to their environment via allele frequency shifts (\citealt{HickeyAndGolding2018}). Ironically, balancing selection is needed to maintain polymorphisms at sufficient frequencies for polygenic adaptation to occur (\citealt{HollingerEtAl2019}), but diploid species are subjected to truncation selection for heterozygosity because they produce many aggregation-prone proteins (main text). That is why the species are diploid to begin with. In this final phase, crossing over is clearly generating genetic diversity that enables rapid environmental adaptation rather than solely facilitating proper chromosome segregation during meiosis, as in the previous phase.
\end{enumerate}
If this four-phase sequence is correct, then meiosis did not originally evolve to support sexual reproduction. Instead, the original function of meiosis was to enable eukaryotes to systematically alternate between haploid and diploid life stages. Meiosis has since been maintained in diploid-dominant species because it promotes polygenic adaptation. Thus, the benefits of meiosis are near-term for haplodiplontic eukaryotes, but long-term for diploid eukaryotes.

\subsection{Asexual reproduction among complex organisms} \label{sec:Asexual}

Two sections ago, I argued that the prokaryotes do not reproduce sexually because they are not complex enough to be subjected to the balancing selection that is required for polygenic adaptation to work. However, some complex organisms that should experience truncation selection for heterozygosity do not reproduce sexually. This section will attempt to explain why.

\cite{HaagAndEbert2004} provides an excellent foundation for explaining asexuality among complex organisms. The authors of the paper argue that asexual species are often found in marginal habitats where species experience many local extinction and recolonization events. The species living in these environments should experience many genetic bottlenecks (founder effect), which would cause substantial genetic drift within the species' populations. As a consequence, these species should experience a decline in genetic diversity over time. The decline in genetic diversity would inflict a large fitness cost on a sexual population because the average individual heterozygosity would decrease with each passing generation. The sexual population would essentially become inbred over time. An asexual population that lacks recombination would not experience this fitness cost because the heterozygosity within each individual organism would be fixed at a constant value, so the asexual population would not experience any inbreeding depression.

The theory presented in \cite{HaagAndEbert2004} works well for asexual hybrids, polyploids, apomictic parthenogens, and any other asexual species with low recombination rates and fixed heterozygosity. However, the theory must be incomplete because some asexual species, such as self-fertilizers and automictic parthenogens, are highly inbred (\citealt{BarriereAndFelix2007}; \citealt{StenoienEtAl2005}; \citealt{Birky1996}). Thus, fixed heterozygosity cannot be the sole advantage of asexual reproduction. However, I do think \cite{HaagAndEbert2004} is on the right track.

Any species subjected to metapopulation dynamics should experience substantial genetic bottlenecks and drift as described in \cite{HaagAndEbert2004}. Polygenic adaptation would be impossible in such circumstances because genetic drift would govern allele frequency changes, making adaptation via allele frequency shifts unlikely. Since sexual reproduction requires polygenic adaptation (see \ref{sec:PolygenicSex}), a species following metapopulation dynamics would not benefit from sexual reproduction. However, sexual reproduction would still be accompanied by fitness costs, such as the two-fold cost of males. Hence, asexual reproduction should be more advantageous than sexual reproduction for species following metapopulation dynamics, as described in \cite{HaagAndEbert2004}. This revised hypothesis should apply to highly inbred asexual species as well as those with fixed heterozygosity.

The importance of heterozygosity has diminished in the revised theory, but it is not irrelevant. Most, but not all, asexual species can be placed in one of two broad categories. There are the highly heterozygous asexual species (hybrids and polyploids) and there are the minimally heterozygous species (self-fertilizers and automicts) (\citealt{JaronEtAl2018}). Both types of asexual species might occupy ephemeral environments that promote metapopulation dynamics, but the harshness of the environments should be considerably different. The highly heterozygous asexual species should be associated with stressful environments that promote high levels of protein aggregation while the minimally heterozygous asexual species should be associated with very mild environments that promote very low levels of protein aggregation. I want to discuss each kind of environment in some detail.

The highly heterozygous asexual species should occupy environments that promote high levels of protein aggregation. As discussed in Section \ref{subsec:Polyploid} of the main text, polyploid and hybrid species are often found in such environments. Asexual species are also found in harsh environments (\citealt{LundmarkAndSaura2006} \citealt{TilquinAndKokko2016}), which is consistent with the metapopulation hypothesis because populations in harsh environments may be restricted local patches that are temporarily hospitable. These species most closely conform to \cite{HaagAndEbert2004}'s original metapopulation hypothesis because metapopulation dynamics should result in a significant loss of heterozygosity in sexual populations. Thus, there are two reasons why asexual species are associated with really harsh environments: (1) asexual species can retain high levels of heterozygosity better than sexual species and (2) genetic drift renders polygenic adaptation ineffectual, thereby removing the main advantage sexual species have over asexual species.

The bdelloid rotifers could provide a good example of highly heterozygous asexual species. Bdelloid rotifers are degenerate tetraploids, and many species survive in environments where the rotifers are routinely exposed to desiccation stress (\citealt{NowellEtAl2018}; \citealt{HurEtAl2009}; \citealt{WelchEtAl2008}). However, \cite{NowellEtAl2018} has found that two limnoterrestrial bdelloid species that are frequently exposed to desiccating conditions possess much higher homologous sequence divergence than two aquatic bdelloid species that are not frequently exposed to desiccation. I think this could be evidence that the limnoterrestrial species are exposed to selection for high levels of heterozygosity. Some form of mitotic recombination seems to be responsible for the homogenization of the aquatic species' genomes (\citealt{NowellEtAl2018}), and these processes should operate on the limnoterrestrial species as well, but the recurring exposure to desiccation stress that limnoterrestrial species experience has probably prevented the limnoterrestrial species from losing as much heterozygosity as the aquatic species. Thus, limnoterrestrial bdelloids probably benefit from asexual reproduction because it helps maintain high levels of heterozygosity as described in \cite{HaagAndEbert2004}. Furthermore, bdelloids constitute 95\% of all rotifers in limnoterrestrial environments, but only 20-30\% of all rotifers in aquatic environments (\citealt{Ricci1987}), which is more evidence that asexual species are more competitive in ephemeral environments than in more permanent environments.

Now consider the opposite extreme of asexual species with very low heterozygosity levels, such as self-fertilizers and automictic parthenogens. According to the main text, such species should only be able to survive if they never experience high levels of protein aggregation. Otherwise, they would be weeded out by truncation selection for heterozygosity. These species should be subject to metapopulation dynamics because the very mild conditions in which they can survive should only last temporarily in localized patches. Genetic drift should inhibit polygenic adaptation in these species, thereby removing the advantage of sexual reproduction as described previously. In addition, these species are subject to low levels of balancing selection, a key requirement for polygenic adaptation (\citealt{HollingerEtAl2019}), which would also render sexual reproduction unbeneficial. Thus, asexual reproduction should be common in species that rarely experience high levels of protein aggregation for two reasons: (1) the absence of balancing selection weakens the efficacy of polygenic adaptation and (2) genetic drift also renders polygenic adaptation ineffectual, thereby removing the main advantage sexual species have over asexual species. \textit{Caenorhabditis elegans} and \textit{Arabidopsis thaliana} could serve as examples of this type of asexual species (\citealt{BarriereAndFelix2007}; \citealt{StenoienEtAl2005}).

I think it would be useful here to make a distinction between stress tolerating species and stress avoiding species. I see this as a generalization of the freeze tolerating and freeze avoiding distinction made in the insect literature (\citealt{Sinclair1999}; \citealt{Bale1996}). Briefly, freeze tolerating species can withstand the formation of ice in their tissues while freeze avoiding species rely on supercooling to prevent the formation of ice in their tissues. Similarly, I think stress tolerating species can withstand the stresses of their environment, but stress avoiding species have evolved mechanisms to limit the damage that physical stresses might inflict upon them. Mechanisms used by stress avoiding species include supercooling, vitrification, and dormancy within seeds or encysted eggs (\citealt{CroweEtAl1998}; \citealt{Siepel1994}).

Thus, self-fertilizers and automictic parthenogens with very low heterozygosity levels may be stress avoiding species, and hybrid parthenogens and asexual polyploids with high heterozygosity levels may be stress tolerating species. \textit{C. elegans}\footnote{\textit{A. thaliana} could also serve as a model for stress avoidance (\citealt{HannahEtAl2006}; \citealt{Reyes-DiazEtAl2006})} and the bdelloid rotifers may serve as models for the two types of species. \textit{C. elegans} dauer larva are able to survive desiccation, but only after they have been preconditioned for 48 hours at 98\% relative humidity. During the preconditioning period, \textit{C. elegans} synthesis trehalose, heat shock proteins (HSPs), and late embryogenesis abundant proteins (LEAs) that prevent protein aggregation from occurring (\citealt{ErkutEtAl2013}). The necessity of a precondition period, the metapopulation dynamics, and the low genetic diversity of \textit{C. elegans} all suggest that it is a typical stress avoiding species (\citealt{TeotonioEtAl2017}; \citealt{ErkutEtAl2013}; \citealt{BarriereAndFelix2007}). In contrast, many bdelloid rotifer individuals can survive a rapid drop in humidity from 95\% to 40\% over one hour (\citealt{CaprioliAndRicci2001}). The high genetic diversity of bdelloid rotifers combined with their ability to survive rapid desiccation would suggest that bdelloid rotifers are typical stress tolerant species.

I have spent much of this section discussing asexual species whose heterozygosity levels are at the high and low end of the spectrum. This is because hybridization, polyploidization, self-fertilization, and automixis are common routes to asexual reproduction. However, asexual species do not have to fall into the two categories I have been discussing. The important issue is whether genetic drift is interfering with polygenic adaptation, which would remove the main advantage of sexual reproduction regardless of the stressfulness of the environment. However, there are extra factors that promote asexual reproduction in stress-tolerating and stress-avoiding species, namely asexual reproduction allows for fixed heterozygosity levels in stress-tolerating species, and the absence of balancing selection in stress-avoiding species provides additional interference inhibiting polygenic adaptation. Therefore, stress-tolerating and stress-avoiding asexual species have extra reinforcements promoting asexual reproduction that other asexual species solely following metapopulation dynamics do not.

I would like to finish this section by briefly mentioning some other asexual species that provide support for the metapopulation hypothesis. The ant species \textit{Mycocepurus smithii} contains sexual and asexual populations. The sexual populations are concentrated in the central part of the species distribution around the Amazon River Basin whereas the asexual species are spread throughout South America and Central America (\citealt{RabelingEtAl2011}). The oribatid mites contain numerous asexual species (\citealt{MaraunEtAl2009}; \citealt{PalmerAndNorton1992}). Asexual oribatid species are rare in arboreal environments, but are common in soil environments(\citealt{DomesEtAl2007}). In fact, asexual oribatid species increase in frequency with soil depth, perhaps because the distribution of food resources becomes increasingly patchy (\citealt{Smelansky2006}). Asexual springtail (Collembola) species also increase in frequency with soil depth (\citealt{ChernovaEtAl2010}). The gecko species \textit{Nactus Arnouxii} contains sexual and asexual populations. The sexual populations are found in Australia, Papua New Guinea, and neighboring islands whereas the asexual populations are found in farther away Pacific islands extending out into Micronesia (\citealt{Moritz1987}). This provides more evidence for the significance of genetic drift in promoting asexual reproduction. Two asexual species of night lizard, \textit{Lepidophyma reticulatum} and \textit{Lepidophyma lavimaculatum}, are found in Costa Rica and Panama, respectively (\citealt{SinclairEtAl2010}). These species are located at the southernmost end of their genus's range, which may again signify a role for genetic drift in promoting asexual reproduction. Asexual non-marine ostracod species are broadly distributed throughout Europe, but sexual species are concentrated near the Mediterranean Sea. \cite{HorneAndMartens1999} argues that asexual species occur in regions frequently disrupted by climate changes and that sexual species occur in regions with a more stable climate.

\subsection{The importance of spores} \label{sec:Spores}

Spores may not seem important, but if the ideas presented in the main text and this appendix are true, then spores were critical to the evolution of complexity, diploidy, and sexual reproduction. That is why I think complexity, diploidy, and sexual reproduction are associated with the sporophyte generation in plants, and with ascocarps and basidiocarps in fungi. I want to briefly discuss the importance of spores in the evolutionary history of life.

The importance of spores is immediately obvious in the evolution of syngamy, a critical step needed for sexual reproduction. As discussed in the beginning of this appendix, many organisms undergo syngamy prior to spore production. This allows the spores to contain a mixed cytoplasm, which could lower protein aggregation rates within the spore. Lowering the rate at which soluble oligomers form is beneficial because it would increase the longevity of the spore and increase the spore's resistance to physical stresses. Thus, syngamy may have evolved because it allowed for the creation of hardier spores.

Spores are great for dispersing offspring over a wide area. In the case of land plants and fungi, spore dispersal is aided by structures that rise above the ground, such as sporophytes, ascocarps, and basidiocarps. These structures must be diploid otherwise the spores they produce would not contain mixed cytoplasm contents. Hence, diploidy is associated with spore dispersing stages in the plant and fungal life-cycles.

It seems inevitable that some spore dispersing life-stages would become larger and more complex over time in order to more effectively scatter their spores. However, I think there is an additional reason why complexity is associated with spore dispersing structures. As discussed in Section \ref{subsec:Complexity} of the main text, there is a problem with the transition from haploid-dominant life-cycles to diploid-dominant life-cycles. I think spore dispersing life-stages helped to overcome this problem.

To recap, truncation selection would prevent a strictly haploid species from producing an abundance of aggregation-prone proteins. Otherwise, such a species would sit at the precipice of a rapid fall in fitness, and probable extinction, should the environment deteriorate. Diploid individuals that might arise in the species would not have any competitive advantage because protein aggregation would not be a significant problem, and there could be advantages to being haploid (Section \ref{subsec:Haploid} of main text). Thus, the evolution of complexity contains a roadblock. Strictly haploid species will not become more complex over time because they cannot produce the aggregation-prone proteins that facilitate complexity, and no diploid species exist because diploidy does not confer any immediate competitive advantage.

However, the previously discussed spore dispersing structures provide a path around the roadblock. The spore dispersing structures are diploid because their offspring benefit from possessing mixed cytoplasm contents, not because the structures themselves benefit from diploidy. This provides enough scope for the spore dispersing structures to evolve the aggregation-prone proteins that facilitate complexity. Hence, some spore dispersing structures have become more complex over time.

I think the roadblock in the evolution of complexity can potentially explain why complexity is often associated with spore dispersing structures. I also think the roadblock can explain why haploid dominant species do not seem to give rise to diploid dominant species directly. Instead, diploid-dominant species seem to typically evolve from species that alternate between haploid and diploid generations, which in turn evolved from haploid-dominant species. This contrasts with the evolution of polyploidy, which occurs in a single generation when diploid parents produce polyploid offspring. The key difference is that the evolution of diploidy is tied to the evolution of complexity, whereas polyploids are typically descended from species that are already complex.

Thus, spores played an important role in the evolutionary history of life. They helped life find a path around the roadblock to the evolution of complexity, and they were the first structures to benefit from diploidy, which favored the evolution of syngamy, a necessary step in the evolution of sexual reproduction. I think this illustrates how complexity, diploidy, sexual reproduction, and spore dispersal are all interconnected.

\subsection{Shifts in truncation selection} \label{sec:Shifts}

Selection for heterozygosity would definitely fall within the parameter space for polygenic adaptation given in \cite{HollingerEtAl2019}. An individual organism's heterozygosity is controlled by thousands of gene loci, and balancing selection maintains multiple alleles at many of the gene loci, at least if the ideas presented in the main text are true. \cite{HollingerEtAl2019} even assumed diminishing returns epistasis for fitness in their models. The truncation selection model presented in the main text may be taken as an extreme form of diminishing returns epistasis. Thus, an individual organism's heterozygosity is a prime candidate for polygenic adaptation via allele frequency shifts.

Some numbers may provide a useful example. Suppose that a population of organisms contains two kinds of alleles at each gene locus subjected to balancing selection. One allele, the major allele, has an average frequency of $p=0.9$. The average frequency of a collection of minor alleles is $q=0.1$. Assuming Hardy-Weinberg equilibrium, then the average individual heterozygosity will be $H=0.18$. Now suppose that the average individual is heterozygous at 1000 gene loci. This must mean that $1000/0.18=5555$ gene loci are subjected to balancing selection. Individual heterozygosity follows a binomial probability distribution because an organism is either heterozygous at a gene locus or it is not. Hence, the standard deviation would be $\sqrt((5555)(0.18)(1-0.18))=29$. If the average allele frequencies were to shift to $p=0.88$ for the major alleles and $q=0.12$ for the minor alleles, then the average heterozygosity would shift to $H=0.21$, and the average individual would be heterozygous at $1167$ gene loci. Thus, very minor allele frequency shifts at many gene loci would cause the average number of heterozygous gene loci, per individual, to move over five standard deviations.

Changes in the average number of heterozygous gene loci per individual could represent an important mode of adaptation for complex organisms. According to the main text, selection for heterozygosity follows a truncation curve in which an individual's fitness dramatically declines below a critical heterozygosity. However, the truncation selection curve could move to either the left or the right, depending on the stressfulness of the environment, as shown in Figure \ref{fig:TruncationSelection}b. When the environment has a high level of stress that promotes protein aggregation (the value of $k_1$ in Figure \ref{fig:TruncationSelection}b is high), then only individuals that are heterozygous at many gene loci will be able to survive in the environment. On the other hand, when the environment is only mildly stressful (the value of $k_1$ in Figure \ref{fig:TruncationSelection}b is low), then individuals with very low heterozygosity levels will be able to survive. Individuals with not heterozygosity at all may be able to survive in very mild environments. Therefore, the individual heterozygosity levels may need to shift when the truncation curve moves to the left and the right.

It is easy to conceive how the shifts in average heterozygosity could occur. Assume that the number of heterozygous gene loci per individual follows a binomial distribution, and that the truncation curve generally stays on the left side of the binomial distribution (otherwise more than half the population would fail to survive each generation). When the truncation curve moves to the right along the binomial distribution, a larger fraction of the population will fail to survive and the average individual heterozygosity for the population will increase. The allele frequencies should shift to accommodate the increase in average individual heterozygosity (see \cite{Wills1978} for an example). On the other hand, only a small fraction of the population will fail to survive when the truncation curve moves to the left, out to the tail-end of the binomial distribution. In this case, a combination of genetic drift and natural selection (assuming that some alleles confer higher fitness than other alleles) would cause allele frequencies to shift in favor of the most common allele, and average individual heterozygosity will decrease in the population (see Equation \ref{eq:Suboptimal} in the main text, and \cite{Wills1978} again). Hence, individual heterozygosity may be constantly changing in natural populations due to a moving truncation curve.

\nociteappendix{*}  
\bibliographyappendix{appendix}
\end{document}